\documentclass[trackchanges]{aastex631}
\usepackage{enumitem,kantlipsum}

\shorttitle{Mode changing in PSR B0844-35 and PSR B1758-29}

\begin{document}

\title{Mode changing in PSR B0844-35 and PSR B1758-29 with enhanced emission at
the profile centers}


\author[0000-0003-1824-4487]{Rahul Basu}
\affiliation{Janusz Gil Institute of Astronomy, University of Zielona G\'ora, ul. Szafrana 2, 65-516 Zielona G\'ora, Poland.}

\author[0000-0002-9142-9835]{Dipanjan Mitra}
\affiliation{National Centre for Radio Astrophysics, Tata Institute of Fundamental Research, Pune 411007, India.}
\affiliation{Janusz Gil Institute of Astronomy, University of Zielona G\'ora, ul. Szafrana 2, 65-516 Zielona G\'ora, Poland.}

\author[0000-0003-1879-1659]{George I. Melikidze}
\affiliation{Janusz Gil Institute of Astronomy, University of Zielona G\'ora, ul. Szafrana 2, 65-516 Zielona G\'ora, Poland.}
\affiliation{Evgeni Kharadze Georgian National Astrophysical Observatory, 0301 Abastumani, Georgia.}

\begin{abstract}
We have studied the single pulse emission from two pulsars, PSR B0844-35 and 
PSR B1758-29, over a wide frequency range of 300-750 MHz using the uGMRT. The
two pulsars have relatively wide profiles with multiple components, that are a 
result of the line of sight traversing near center of the emission beam. In 
both pulsars the single pulse sequences show the presence of two distinct 
emission states, where the profiles become much brighter at the center with
prominent core components during one of the modes, while in the other mode the 
single pulses show odd-even subpulse drifting with periodicity around 2$P$, $P$
being the rotation period of the pulsar. The centrally bright mode was seen for
10 percent of the observing duration in PSR B0844-35, which usually lasted for
short durations of around 10 pulses, but had two longer sequences of around 100
pulses. On the contrary the centrally bright mode was dominant in PSR B1758-29 
and was seen for around 60 percent of the observing duration. PSR B1758-29 also
showed period amplitude modulations of 60-70$P$ in both modes. The mode 
changing in these two pulsars facilitates investigation of the sparking process
in the inner acceleration region, dominated by non-dipolar magnetic fields. The
change in the surface magnetic field configurations likely results in the 
emission mode change.
\end{abstract}


\section{Introduction}
The single pulse emission from pulsars show several distinct phenomena that
include subpulse drifting, seen as systematic variations of the subpulses 
within the emission window resembling regular drift bands \citep{WES06,BMM16,
BMM19,SWS23}; periodic/quasi-periodic modulations where transitions take place 
between emission states with different intensity levels \citep{MR17,BMM17,
BMM20a}; nulling or disappearance of the single pulse emission for varying 
durations \citep{R76,WMJ07,GJK12,BMM17,BMM20a}; and mode changing between two 
or more distinct emission states at regular intervals \citep{WMJ07,GBM21,
BMM21}. These phenomena are symptomatic of the physical processes in the pulsar
magnetosphere that generate the outflowing plasma responsible for the observed 
radio emission. The outflowing plasma is generated as a non-stationary flow due
to sparking discharges in an inner acceleration region (IAR) above the polar 
cap \citep{S71,RS75}. The IAR is characterised by highly non-dipolar magnetic 
fields with large electric potential difference along the magnetic field lines 
\citep{G17,AM19,SG20,PM20}. 

A systematic approach towards understanding the sparking process has been 
developed in recent years where the IAR resembles a Partially Screened Gap 
(PSG) with a steady outflow of positively charged ions partially screening the 
gap potential \citep{GMG03}. The sparks in the PSG develop due to cascading 
electron-positron pair production in the large potential difference, where the 
positrons are accelerated to large energies, Lorentz factor $\gamma\sim10^6$, 
away from the star. The electrons are accelerated downwards and heat the 
surface to critical temperatures of 10$^6$ K, which causes the positively 
charged ions to freely escape from the surface and screen the gap potential 
\citep{CR80,J86}. The PSG provides the only known physical model for forming a 
stable two dimensional system of sparks across the polar cap with well defined 
perpendicular spark size, that is obtained from the thermal regulation 
criterion of the IAR \citep{MBM20,BMM20b,BMM22b}. The sparks have typical 
lifetimes of several tens of microseconds which is the time required to heat 
the surface to critical temperature. Effective heat regulation in the PSG 
requires the sparks to be formed in a tightly packed configuration in the IAR.

During sparking the plasma lag behind the rotation of the star, as the charge 
density is less than the Goldreich-Julian density, shifting the location of 
maximum heating. The entire polar cap rotates along with the pulsar around the 
rotation axis, while all the sparks formed inside the IAR above the polar cap 
also move around the rotation axis, but with a lower speed than the star, such 
that in the frame of the polar cap they shift opposite to the direction of 
rotation. Additionally, the sparks cannot be formed in the closed field line 
region, hence, the boundary of the polar cap also constrains the next spark to 
be formed along the boundary. In the plane of the polar cap with the axis 
located at the center, along the outward normal to the plane, and the equator 
defined as the diameter along the rotation motion cutting across the polar cap,
the sparks in the northern half are formed shifted in the clockwise direction 
around this axis, while those in the southern half are formed shifted in the 
anti-clockwise direction. At the equatorial boundary the sparks above and below
shifts in opposite directions creating an opening that either grows (leading 
side) or shrinks (trailing side) with time. Sparking is triggered only when the
space is large enough for the spark to form. Near the center of the polar cap, 
around the axis, due to lack of available space, the sparks do not shift in 
either direction but appears at the same location at regular intervals 
\citep[see animations in][for visualization of the sparking evolution in 
different polar cap configurations]{BMM22b}. This sets up a steady drift like 
pattern in the outflowing plasma with a stationary center. 

The primary plasma generated in the sparks at the IAR continues to produce 
secondary pair particles with Lorentz factor $\gamma\sim10^2$ outside the gap, 
making up the outflowing plasma clouds. The radio emission is generated due to 
non-linear instabilities developing in these outflowing plasma clouds that 
preserve the imprint of the sparks in the observed emission \citep{MP80,MGP00,
GLM04,M17,LMM18,RMM20,RMM22a,RMM22b}. The drift pattern in the spark dynamics 
is observed as subpulse drifting in the single pulse emission \citep{BMM20b}. 
Hence, it is possible to find suitable estimates for the physical parameters of
the IAR, like the average screening factor of the electric potential difference
in a PSG, the structure of surface non-dipolar magnetic fields, etc., from the
pulsar subpulse drifting measurements \citep[see][for details]{BMM23}. The 
evolution of the sparking system in the IAR is also evident in the average 
profiles of the pulsar population. The pulsar radio emission beam has been 
described using the core-cone model, where the average emission has the form of
a central core and two rings of nested conal emission, namely, the inner and 
outer cones \citep{ET_R90,ET_R93,MD99}. The average profile shape depends on 
the line of sight (LOS) traverse across the beam, with central LOS cuts giving 
rise to core-cone Triple (T) and core-double cone multiple (M) types, while the
shape resembles conal Quadruple ($_c$Q), conal Triple ($_c$T), double (D) and 
conal single (S$_d$) as the LOS moves further away from the center. It has been
observationally confirmed that no drifting behaviour is seen in the central 
core component and only appears in the surrounding cones \citep{BMM19,BPM19,
BLK20}, as expected from the evolution of sparking pattern in the PSG.

The other modulation features seen in the single pulse sequence, like mode
changing, nulling and periodic modulations, are associated with change in the
radio emission behaviour across the entire profile window, usually over 
timescales of less than a rotation period. It is likely that such variations 
are a result of small scale changes in the surface non-dipolar field 
configuration, with the global dipolar magnetic field remaining unchanged. The 
physical mechanism that can cause such transitions between the different 
magnetic field configurations is still not well understood. There are proposals
for introducing perturbations by the Hall drift and thermoelectrically driven 
magnetic field oscillations that can account for such fast changes 
\citep{GBM21}. 

The pulsars showing multiple modes in their single pulse emission, especially 
where one or more modes exhibit subpulse drifting behaviour, are unique 
laboratories to understand the surface parameters of the PSG as well as 
possible modifications during mode changing. For a long time only a handful of 
pulsars were known to exhibit mode changing associated with subpulse drifting,
with prominent examples being, PSR B0031-07 \citep{HTT70,VJ97,SMK05,MBT17}, PSR
B0809+74 \citep{vLKR02,HSW13,BLK23}, PSR B0943+10 \citep{SIR98,BMR11}, PSR 
B1237+25 \citep{B70b,SR05}, PSR B1918+19 \citep{HW87,RWB13}, PSR B1944+17 
\citep{DCH86,KR10}, PSR B2303+30 \citep{RWR05}, and PSR B2319+60 \citep{WF81,
RBMM21}. In recent years dedicated studies of the single pulse emission from 
pulsars using instruments like GMRT, MWA, FAST, etc., have found several other 
examples of subpulse drifting associated with mode changing, e.g. PSR 
J0026-1955 \citep{MBS22,JMC23}, PSR B0820+02 \citep{ZXS23}, PSR J1110-5637 
\citep{DSL22}, PSR J1727-2739 \citep{WWY16,BMM21,RWY22}, PSR B1819-22 
\citep{BM18b,JCB22}, PSR B2003-08 \citep{BPM19}. In the above list only three 
pulsars, PSR B1237+25, PSR B2003-08 and PSR B2319+60 have $_c$Q/M profile types
signifying the central LOS traverse across the emission beam. The central LOS 
traverse provides a wider window into the polar cap region to provide better 
constraints on the IAR parameters. 

In this work we investigate the the single pulse behaviour of two pulsars PSR 
B0844-35 and PSR B1758-29 with multiple components in their average profiles. 
Both pulsars have been reported to show emission mode changing as well as 
subpulse drifting in their single pulse sequence \citep{WMJ07,BMM16,BMM21}. We 
have observed the long sequences of single pulses from these two sources over a
wide frequency range to explore the connection between these two phenomena. 
This will enable the characterisation of the surface properties in the IAR, 
including the PSG parameters and the surface magnetic field configuration. In 
section 2 we report the observational details. Section 3 presents the single 
pulse and average profile behaviour of PSR B0844-35 while the equivalent 
studies for PSR B1758-29 is reported in section 4. Finally, a discussion of the
physical properties in these two systems are carried out in section 5, 
including the changes of the PSG parameters required for the different emission
states. However, a detailed modelling of the surface properties based on the 
observations requires a separate study dedicated to this problem.

\begin{deluxetable}{ccccccccccc}
\tablecaption{Observing Details \label{tab:obs}}
\tablewidth{0pt}
\tablehead{
 \colhead{PSR} & \colhead{$P$} & \colhead{DM} & \colhead{$\dot{E}$} & \colhead{Date} & \colhead{Obs. Mode} & \colhead{Frequency} & \colhead{Npulse} & \colhead{Mode} & \colhead{\%} & \colhead{Avg. Len} \\
    & \colhead{(s)} & \colhead{(cm$^{-3}$~pc)} & \colhead{(erg~s$^{-1}$)} &   &   & \colhead{(MHz)} & \colhead{($P$)} &   &   & \colhead{($P$)} }
\startdata
 B0844-35 & 1.116 & 94.16 & 4.56$\times10^{31}$ & 27/12/2018 & Phased-Array & 300-500 & 1860 & A & 86.8 & 161.5 \\  
    &   &   &   &   &   &   &   & B & 13.2 & 27.2 \\  
    &   &   &   & 17/01/2020 & Dual Sub-array & 300-500, 550-750 & 3780 & A & 92.2 & 388.0 \\  
    &   &   &   &   &   &   &   & B & 7.8 & 36.2 \\  
    &   &   &   &   &   &   &   &   &   &   \\  
 B1758-29 & 1.082 & 125.61 & 1.03$\times10^{32}$ & 18/03/2019 & Phased-Array & 300-500 & 2100 & B & 66.0 & 346.8 \\  
    &   &   &   &   &   &   &   & Q & 34.0 & 178.3 \\  
    &   &   &   & 10/02/2020 & Dual Sub-array & 550-750 & 5261 & B & 57.8 & 190.7 \\  
    &   &   &   &   &   &   &   & Q & 42.1 & 158.6 \\  
\enddata
\tablecomments{The pulsar parameters $P$, DM and $\dot{E}$ were obtained from
ATNF Pulsar Catalogue https://www.atnf.csiro.au/research/pulsar/psrcat/ 
\citep{MHT05}.}
\end{deluxetable}

\section{Observation and Analysis}\label{sec:obs}
We have carried out single pulse observations of PSR B0844-35 and PSR B1758-29 
using the wideband receivers of the upgraded Giant Meterwave Radio Telescope
\citep[u-GMRT,][]{GAK17}. The GMRT consists of 30 antennas, with 14 antennas
located within a central square kilometer area, while the other 16 antennas are
spread out in a Y-shaped array with maximum distance of 27 kilometers. The 
high time resolution observations for pulsar studies are carried out by 
combining the signals from several antennas into a Phased-Array, ensuring 
higher detection sensitivity for the single pulses. It is also possible to
form two or more Sub-Arrays from different antenna configurations, where each
Sub-Array can operate at different frequency bands. We have used different 
observing setups to measure the single pulse emission from both pulsars. 
Initially, the pulsars were observed in the Phased-Array mode comprising of 22 
antennas, the 14 central square antennas and first 2/3 arm antennas, over a 200 
MHz frequency range between 300-500 MHz, for relatively shorter durations. The 
single pulses were clearly detected during these studies which prompted 
subsequent longer observations in the dual Sub-Array mode at two different 
frequency ranges of 200 MHz bandwidth each, resulting in wider, near 
continuous, frequency coverage of 400 MHz between 300 MHz and 750 MHz. The 
first Sub-Array was setup between frequencies of 300-500 MHz, and consisted of 
14 antennas equally divided between the central-square and arm antennas. The 
second Sub-Array operated between the frequency range of 550-750 MHz and 
consisted of 10 antennas, 7 from the central square and one from each arm. A
strong nearby point-like source was observed before the start of the 
observations to correct for the phase deviations in each antenna response, 
thereby maximizing the detection sensitivity of the single pulses after the 
measurements from each antenna were co-added, a process known as phasing of the 
antennas. Table \ref{tab:obs} lists the basic physical properties of each 
pulsar and the details of the observing setup including the dates of 
observations, the observing mode and total single pulses recorded during each 
observation, and emission mode statistics discussed in the next section. PSR 
B0844-35 was observed in the Phased-Array mode on 27 December, 2018 for a total
of 1860 single pulses and later another 3780 single pulses were observed in the
dual Sub-Array mode on 17 January, 2020. PSR B1758-29 was observed in the 
Phased-Array mode on 18 March, 2019 for a total of 2100 single pulses. On 10 
February, 2020 the pulsar was once again observed in the dual Sub-Array mode 
for 5271 single pulses, with a 10 minute phasing interval in between. The lower
frequency observations on this day, between 300-500 MHz, were affected by radio
frequency interference (RFI) and could not be used for the single pulse 
studies. 

The polarization properties of the incident signals from the pulsars were 
recorded in the auto and cross-correlated form. The 200 MHz frequency bandwidth
was divided into 2048 channels. A number of intermediate analysis were carried 
out to obtain a properly calibrated polarized single pulse sequence from the 
time series signals recorded during each observing session. The polarized 
signals were suitably calibrated and converted into the well known four Stokes 
parameters (I, Q, U, V), for each spectral channel \citep[see][for 
details]{MBM16}. Further analysis involved removal of RFI affected signals from
the time sequence as well as narrow frequency channels \citep{MBM16,BMM16}. The
delay of the pulse emission across the frequency band due to dispersion in the 
interstellar medium was corrected using the known dispersion measure of each 
pulsar (see Table \ref{tab:obs}), and averaged over five separate sub-bands 
each of 30 MHz bandwidths centered around 315 MHz, 345 MHz, 397 MHz, 433 MHz 
and 468 MHz, respectively, for the lower frequency band, and similarly into 
five sub-bands at the upper frequency band centered around 576 MHz, 610 MHz, 
643 MHz, 676 MHz and 709 MHz, respectively, to study the frequency evolution of
the emission properties. The sub-bands are not equispaced due to the presence 
of narrow band RFI that were discarded. The I-Stokes parameter of the single 
pulse sequence was re-sampled to form a three-dimensional pulse stack and 
represented in plots with the rotation phase along the x-axis, the pulse number
along the y-axis and the intensity level signified by a colour scale 
\citep{BMM16}. The single pulse analysis pertaining to subpulse drifting, 
nulling, periodic modulation and emission mode changing were carried out using 
the plots of the pulse stacks. On the other hand the average profiles from the 
remaining three Stokes parameters were used to estimate the linear and circular
polarization levels across the emission window, while their single pulse 
sequences were used for detecting the presence of orthogonal polarization modes
and estimating the polarization position angle (PPA) for emission height and 
LOS geometry studies. Below we report the results of the single pulse analysis 
of each pulsar.

\section{Radio emission features in PSR B0844-35}\label{sec:B0844-35}

\subsection{Emission States}
The single pulse emission properties of PSR B0844-35 was initially reported in 
\citet{WMJ07} based on observations at 1.5 GHz using the Parkes 64-m radio 
telescope. The pulsar was also part of the Meterwavelength Polarimetric 
Emission Survey \citep[MSPES,][]{MBM16}, where polarized single pulses were 
observed at 325 MHz and 610 MHz using the GMRT, and the mode changing behaviour
was studied in \citet{BMM21}. The pulsar has four components in the average 
profile of the primary emission state termed as mode A, with the second 
component showing a wider composite structure, and has been tentatively 
classified as a conal only profile of $_c$Q type. The earlier studies also 
reported the presence of a short-duration mode B, lasting between 10-20 
periods at a time, with the emission becoming brighter for the second component
while the remaining components were much weaker during this mode.

\begin{deluxetable}{ccccccc}
\tablecaption{Mode Sequence of PSR B0844-35 \label{tab:seqB0844}}
\tablewidth{0pt}
\tablehead{
 \multicolumn{3}{c}{\underline{27/12/2018}} &  & \multicolumn{3}{c}{\underline{17/01/2020}} \\
 \colhead{Pulse Range} & \colhead{Mode} & \colhead{Mode Length} &  & \colhead{Pulse Range} & \colhead{Mode} & \colhead{Mode Length} \\
 \colhead{($P$)} &   & \colhead{($P$)} &   & \colhead{($P$)} &   & \colhead{($P$)}}
\startdata
    1 - 108  & A & 108 &   &    1 - 323  & A &  323 \\
  109 - 123  & B &  15 &   &  324 - 461  & B &  138 \\
  124 - 269  & A & 146 &   &  462 - 578  & A &  117 \\
  270 - 372  & B & 103 &   &  577 - 591  & B &   15 \\
  373 - 474  & A & 102 &   &  592 - 1457 & A &  866 \\
  475 - 494  & B &  20 &   & 1458 - 1482 & B &   25 \\
  495 - 727  & A & 233 &   & 1483 - 1791 & A &  309 \\
  728 - 735  & B &   8 &   & 1792 - 1813 & B &   22 \\
  736 - 773  & A &  38 &   & 1814 - 2158 & A &  345 \\
  774 - 807  & B &  34 &   & 2159 - 2166 & B &    8 \\
  808 - 914  & A & 107 &   & 2167 - 2410 & A &  244 \\
  915 - 934  & B &  10 &   & 2411 - 2430 & B &   20 \\
  935 - 1011 & A &  77 &   & 2431 - 2567 & A &  137 \\
 1012 - 1022 & B &  11 &   & 2568 - 2599 & B &   32 \\
 1023 - 1565 & A & 543 &   & 2600 - 2713 & A &  114 \\
 1566 - 1581 & B &  16 &   & 2714 - 2743 & B &   30 \\
 1582 - 1605 & A &  24 &   & 2744 - 3780 & A & 1037 \\
 1606 - 1623 & B &  18 &   &   &   &   \\
 1624 - 1860 & A & 237 &   &   &   &   \\
\enddata
\end{deluxetable}

\begin{figure}
\epsscale{0.75}
\plottwo{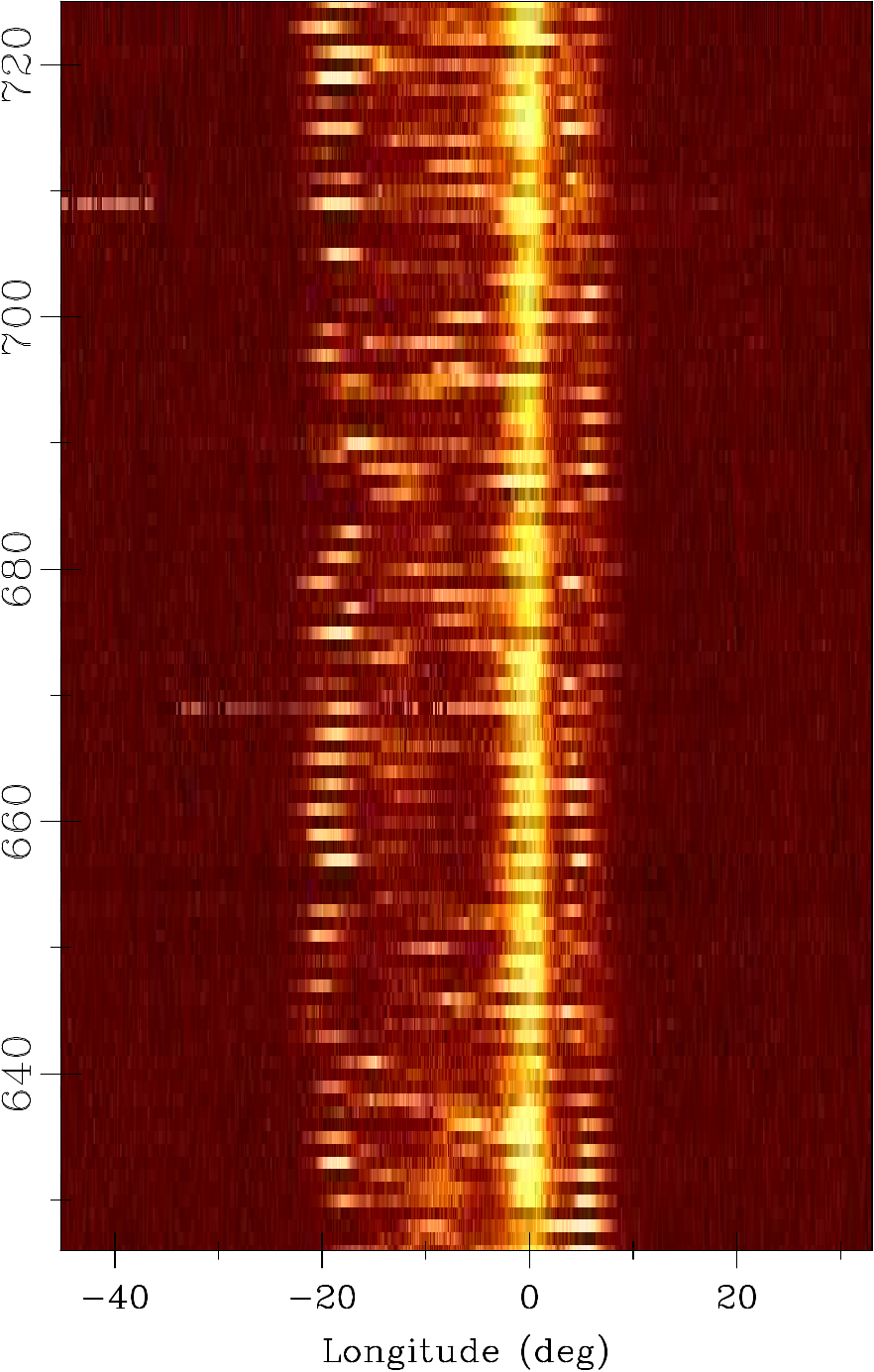}{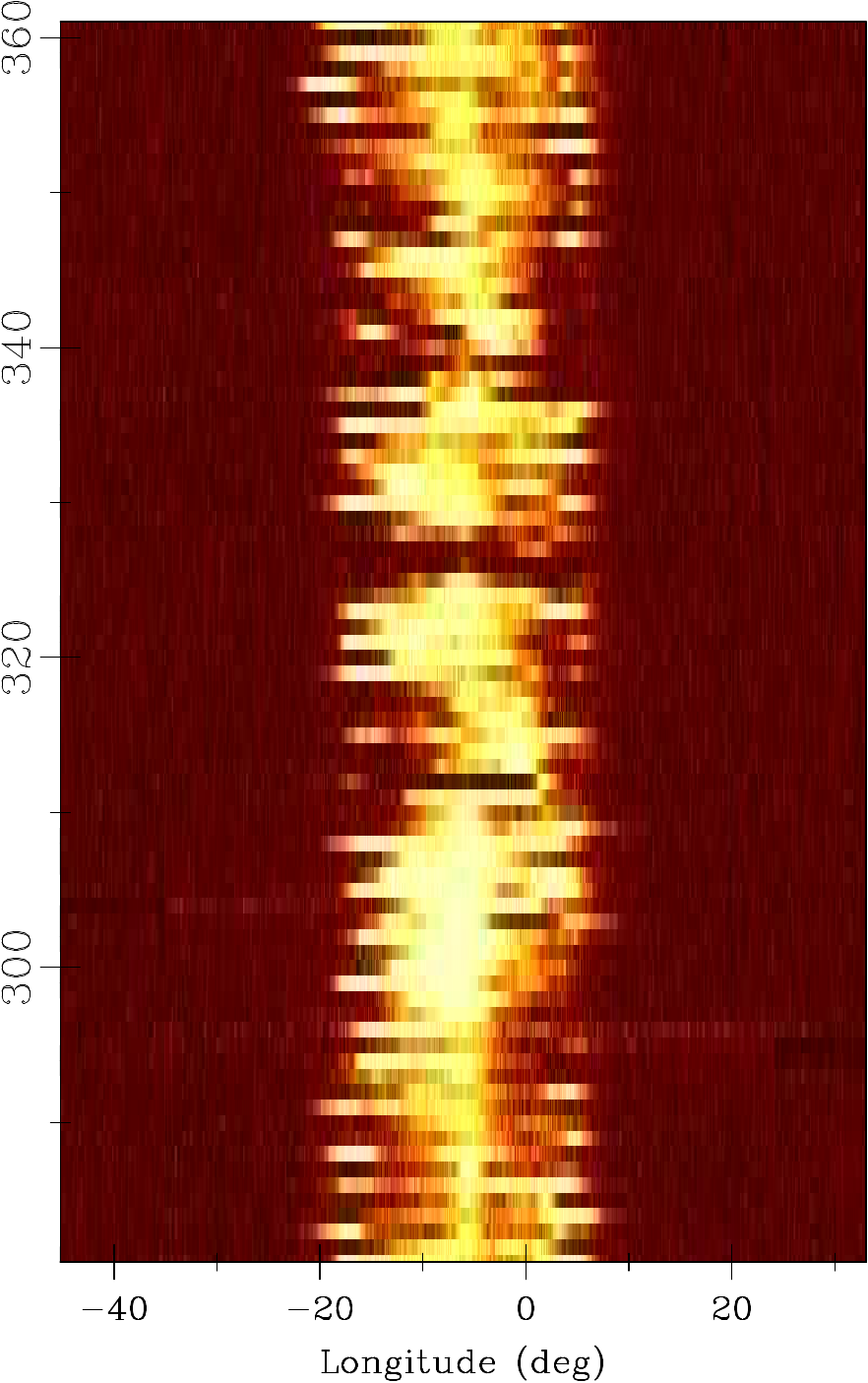}
\caption{The left panel shows the single pulse sequence corresponding to mode A
of PSR B0844-35, observed on 27 January, 2018, and represents the pulse range
between 625 to 725 from the start of the observing session averaged over the 
300-500 MHz frequency band. The right panel shows the 300-500 MHz frequency 
averaged pulse sequence also observed on 27 January, 2018, between pulses 280 
and 360, belonging to mode B.
\label{fig:singlB0844}}
\end{figure}

We identified the mode changing behaviour in this pulsar over the different 
observing sessions by careful visual inspection of the single pulse sequences 
in the pulse stacks and detected the two emission modes. The mode changing was 
simultaneously seen across the entire frequency range, between 300 MHz and 750 
MHz, highlighting its broadband nature. Table \ref{tab:seqB0844} shows in 
detail the sequence of the emission modes during each observing session, and 
the average statistics of the modes are reported in Table \ref{tab:obs}. Mode A
was seen for around 90 percent of the observing duration and lasted for several 
hundred periods at a time. In the remaining 10 percent of time the pulsar 
switched to Mode B, which typically lasted for 20-30 periods. In addition to 
the short duration modes we also detected two instances, out of 17 transitions 
to mode B, when the mode was seen for longer durations in excess of hundred 
rotation periods, once during each observing session.

\begin{figure}
\plottwo{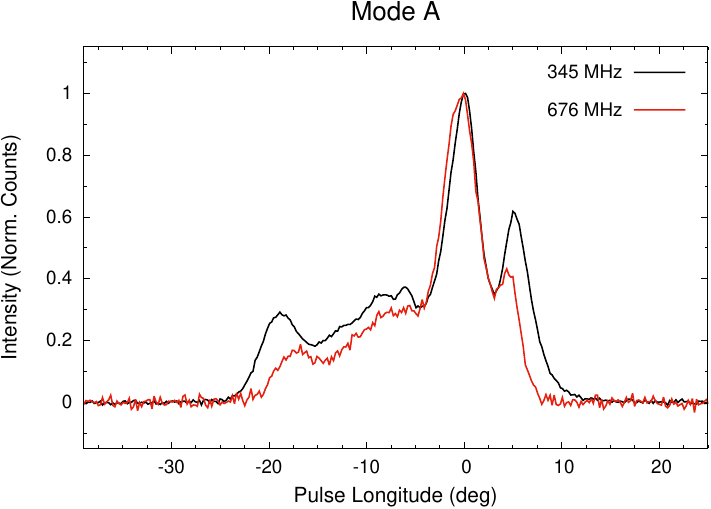}{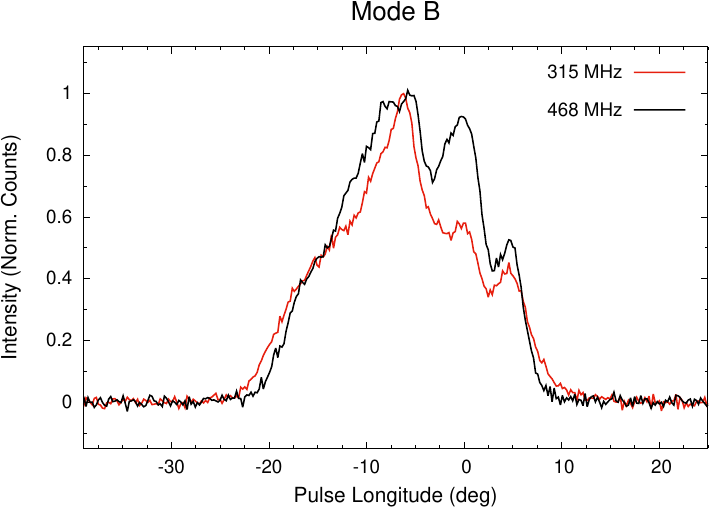}
\caption{The average profiles of PSR B0844-35 during Mode A (left panel) and 
Mode B (right panel). The profiles at two separated frequencies are shown in 
each panel to highlight the frequency evolution of the modes.
\label{fig:profB0844}}
\end{figure}

\begin{figure}
\epsscale{0.6}
\plotone{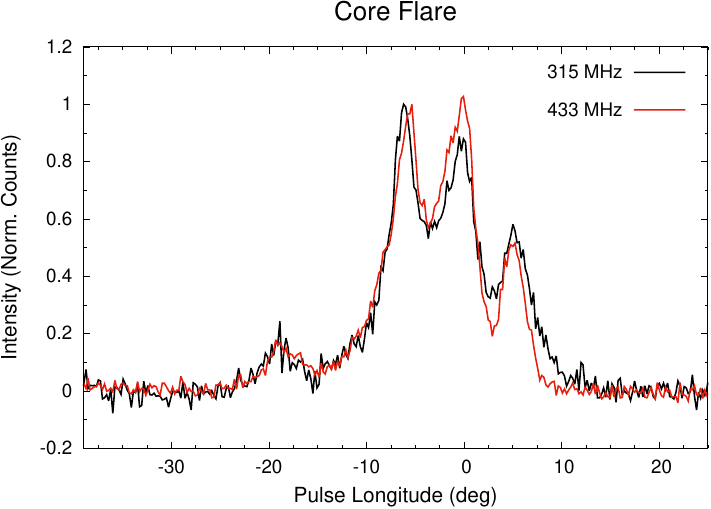}
\caption{The average profile from selected pulses where there is flaring in the
core emission.
\label{fig:flareB0844}}
\end{figure}

\begin{deluxetable}{ccccccc}
\centerwidetable
\tablecaption{Average Profile properties in the emission modes of PSR B0844-35 \label{tab:profB0844}}
\tablewidth{0pt}
\tablehead{
 \colhead{Mode} & \colhead{Frequency} & \colhead{$W_{5\sigma}$} & \colhead{$W_{10}$} & \colhead{$W_{50}$} & \colhead{$W_{sep}^{in}$} & \colhead{$W_{sep}^{out}$} \\
    & \colhead{(MHz)} & \colhead{(\degr)} & \colhead{(\degr)} & \colhead{(\degr)} & \colhead{(\degr)} & \colhead{(\degr)}}
\startdata
     & 315 & 37.0$\pm$0.5 & 34.0$\pm$0.5 & 28.9$\pm$0.5 & 8.5$\pm$0.5 & 24.3$\pm$0.5 \\
     & 345 & 36.8$\pm$0.5 & 32.9$\pm$0.5 & 28.2$\pm$0.5 & 8.4$\pm$0.5 & 23.9$\pm$0.5 \\
     & 397 & 35.7$\pm$0.5 & 31.3$\pm$0.5 & 27.4$\pm$0.5 & 8.2$\pm$0.5 & 23.8$\pm$0.5 \\
     & 433 & 33.4$\pm$0.5 & 30.8$\pm$0.5 & 27.1$\pm$0.5 & 8.2$\pm$0.5 & 23.5$\pm$0.5 \\
  A  & 468 & 32.1$\pm$0.5 & 30.2$\pm$0.5 & 26.7$\pm$0.5 & 8.2$\pm$0.5 & 23.3$\pm$0.5 \\
     & 576 & 28.5$\pm$0.5 & --- & 25.7$\pm$0.5 & --- & 22.1$\pm$0.5 \\
     & 610 & 29.8$\pm$0.5 & --- & 25.6$\pm$0.5 & --- & 21.8$\pm$0.5 \\
     & 643 & 29.2$\pm$0.5 & --- & 25.9$\pm$0.5 & --- & 21.4$\pm$0.5 \\
     & 676 & 28.3$\pm$0.5 & --- & 25.5$\pm$0.5 & --- & 21.5$\pm$0.5 \\
     & 709 & 27.9$\pm$0.5 & --- & 25.5$\pm$0.5 & --- & 21.2$\pm$0.5 \\
     &     &    &   &   &   &   \\
     & 315 & 33.6$\pm$0.5 & --- & 20.9$\pm$0.5 & --- & --- \\
     & 345 & 32.8$\pm$0.5 & 30.6$\pm$0.5 & 20.6$\pm$0.5 & --- & --- \\
     & 397 & 31.7$\pm$0.5 & 28.6$\pm$0.5 & 20.2$\pm$0.5 & --- & --- \\
     & 433 & 31.1$\pm$0.5 & 28.3$\pm$0.5 & 20.0$\pm$0.5 & --- & --- \\
  B  & 468 & 29.8$\pm$0.5 & --- & 20.5$\pm$0.5 & --- & --- \\
     & 576 & 25.2$\pm$0.5 & --- & 20.3$\pm$0.5 & --- & --- \\
     & 610 & 24.9$\pm$0.5 & --- & 19.0$\pm$0.5 & --- & --- \\
     & 643 & 24.9$\pm$0.5 & --- & 19.0$\pm$0.5 & --- & --- \\
     & 676 & 25.6$\pm$0.5 & --- & 18.6$\pm$0.5 & --- & --- \\
     & 709 & 23.9$\pm$0.5 & --- & 18.1$\pm$0.5 & --- & --- \\
\enddata
\end{deluxetable}

The left panel in Fig.~\ref{fig:singlB0844} shows a pulse sequence during mode 
A while the right panel shows the long sequence of mode B during the first 
observing session. Fig.~\ref{fig:profB0844} shows the average profiles of the 
two modes at two widely spaced frequency bands. The single pulse emission 
during mode A shows the presence of odd-even drifting pattern (see section 
\ref{sec:B0844drift}). The wider second component shows bifurcation at the 
lower frequencies (see Fig.~\ref{fig:profB0844}, left panel), which suggest the 
appearance of the central core component which is otherwise masked by the 
leading inner cone. At higher frequencies there is no clear separation between 
the inner cone and the core. The single pulse emission also shows certain 
instances where the core appears to flare up, similar to certain pulsars with
partial cone profiles \citep{MR11} and central flaring seen in a few cases with
D type profiles \citep{YR12}. The estimated average profile from 41 pulses on 
the first observing session, where the core emission flares up is shown in 
Fig.~\ref{fig:flareB0844}. The leading inner cone is much weaker during these 
pulses, thereby the core becomes clearly visible (see Fig.~\ref{fig:flareB0844}
between longitude range -10\degr~and -5\degr). The core also appears to be more
prominent at the low frequency emission of mode B where the conal components 
are suppressed (see Fig.~\ref{fig:profB0844}, right panel, profile at 345 MHz 
in red). However, due to the differential spectral evolution of the core and 
the conal components, with the core having a steeper spectral index than the 
inner cone \citep{BMM21,BMM22a}, a clear transition in the mode B profile shape
takes place above 450 MHz, with the inner cone becoming comparable and hiding 
the core. 

The evolution of the profile widths as a function of frequency during the two
emission states is reported in Table \ref{tab:profB0844}, including the width
estimated at 5 times the noise rms level ($W_{5\sigma}$), the width at 10 
percent ($W_{10}$) and 50 percent ($W_{50}$) of the peak intensity of the outer
most components on either side of the profile, and the separation between the
inner ($W_{sep}^{in}$) and outer ($W_{sep}^{out}$) conal pairs, where the peak 
position of each component is identified by locating the centroid. The effect 
of radius to frequency mapping along dipolar magnetic field, i.e. lower 
frequencies originating higher up the pulsar magnetosphere, is evident in the 
profile widths of mode A as well as the separation between the outer conal 
components \citep{ET_MR02}. However, the separation between the inner cones do 
not show any significant change with varying frequency, which also agrees with 
earlier measurements in a number of other pulsars. In case of mode B the conal 
components become relatively weaker and the separation between the conal pairs 
could not be measured. The evolution of the profile shape due to the relative 
spectral difference between the core and conal emission also become evident in 
the $W_{50}$ measurements, which shows an increase between the 433 MHz and 468 
MHz profile widths.

In the 2 hour observations of \citet{WMJ07}, the authors did not detect the
presence of nulling in this pulsar. We did not find any systematic nulling 
behaviour in this pulsar, even during the more sensitive observations on 27 
January, 2018, at the 300-500 MHz band, which agrees with the earlier results.
There were 16 pulses where the emission was below the detection limit which 
puts the upper limit for the nulling fraction to be below 1 percent. However, 
these were likely a result of the stochastic variations expected from the 
emission mechanism \citep{RMM20,RMM22a,RMM22b,MMB23a}, and not associated with 
changes in the polar cap configuration, that we believe is the likely cause of 
nulling in pulsars \citep{GBM21}.

\subsection{Subpulse Drifting}\label{sec:B0844drift}
\begin{figure}
\plottwo{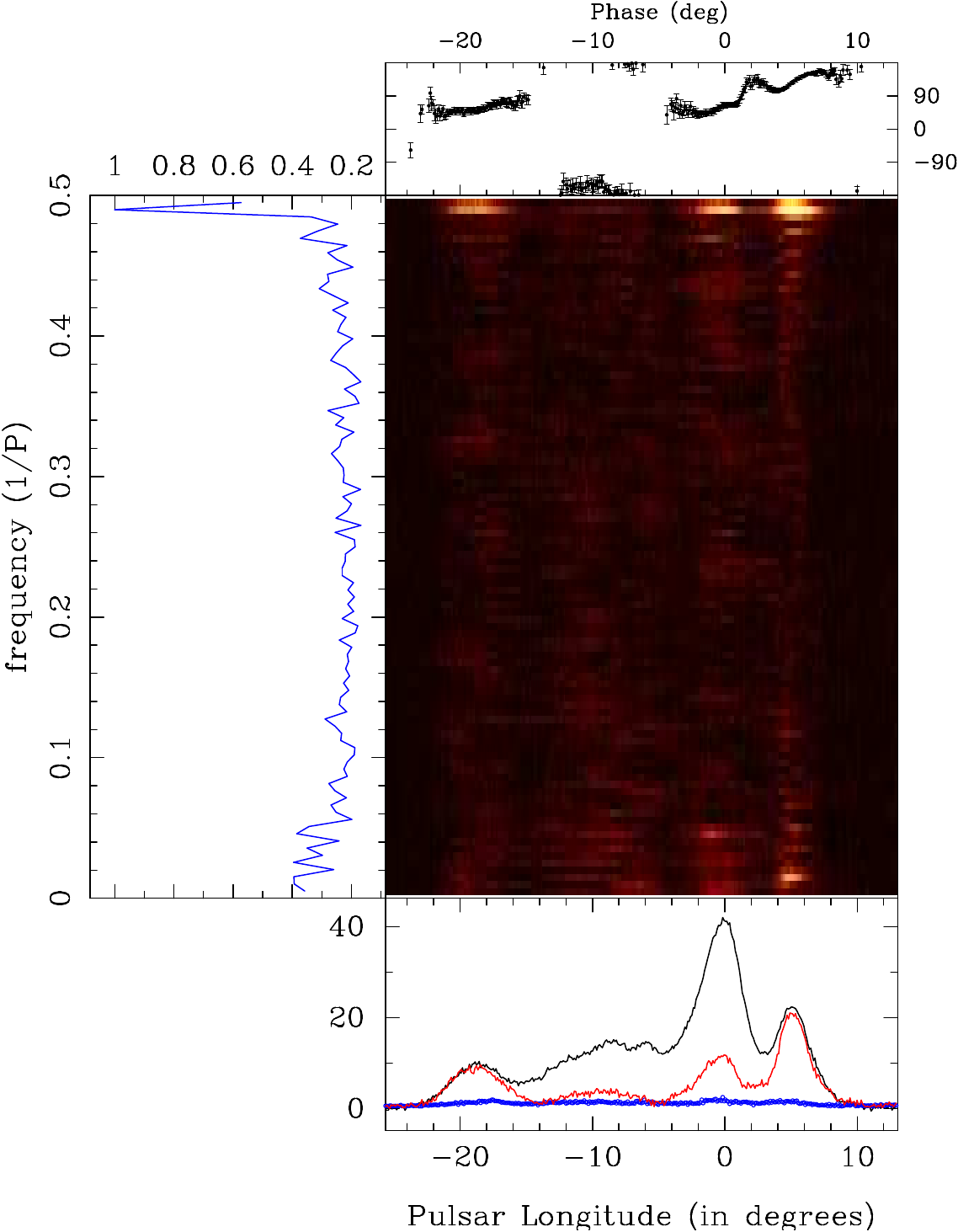}{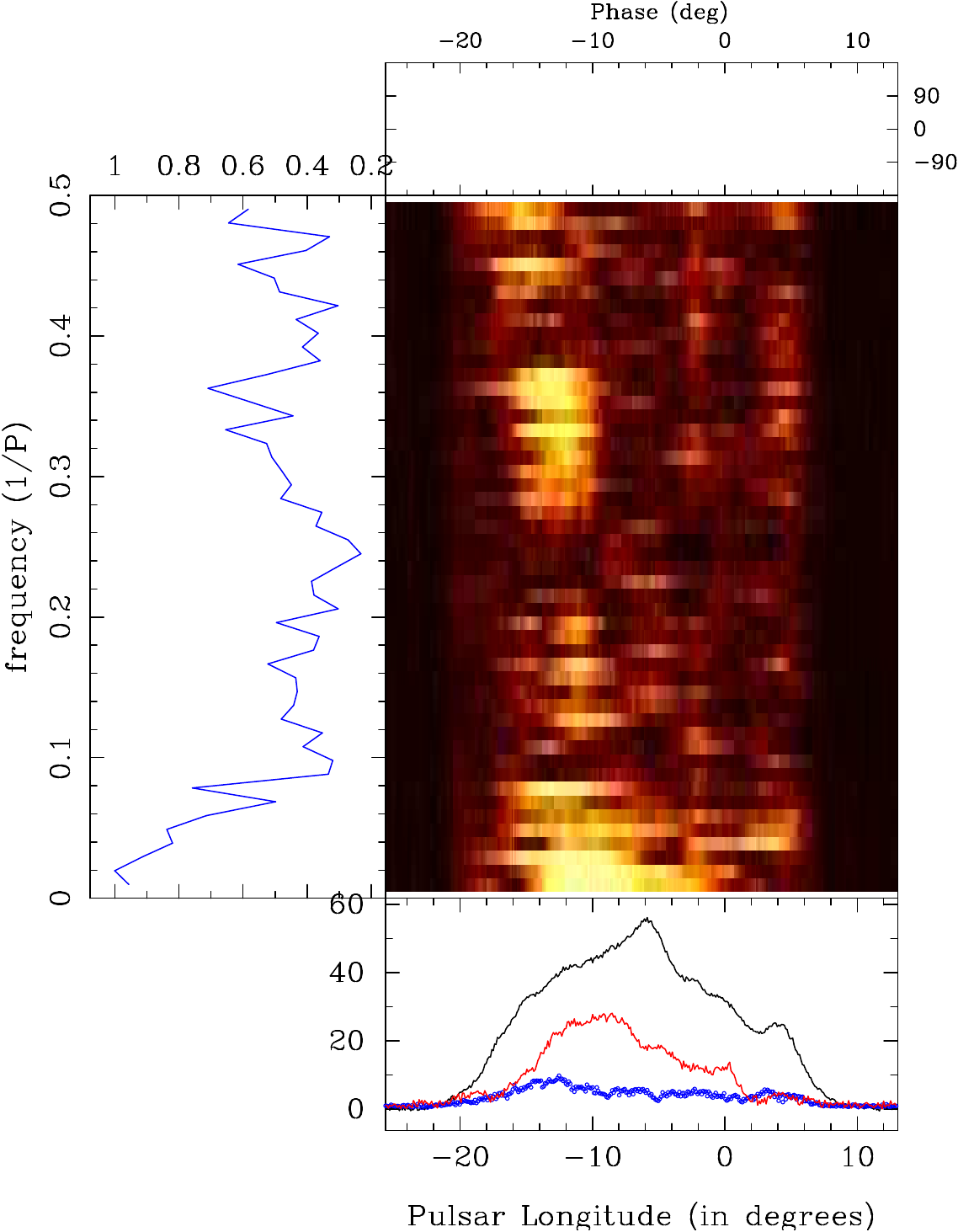}
\caption{Longitude resolved fluctuation spectra (LRFS) to estimate the drifting
behaviour in the two modes of PSR B0844-35. The left panel shows the LRFS for
a pulse sequence of 180 periods in mode A and the right panel shows the LRFS of
mode B for 102 pulses. A peak frequency due to subpulse drifting with 
periodicity $\sim2P$ is seen during mode A, while mode B do not show any clear 
periodic behaviour. 
\label{fig:lrfsB0844}}
\end{figure}

The presence of subpulse drifting in PSR B0844-35 was detected in drifting 
studies of the MSPES sample \citep{BMM16,BMM19}. However, these studies did not
explore the drifting properties of the individual modes. The periodic 
variations associated with the subpulse drifting behaviour is measured using 
the mathematical technique of fast Fourier transforms (FFT). The FFT is carried
out for a sequence of numbers signifying the emission intensity of consecutive 
single pulses at a given longitude in the emission window. This process is 
repeated for all longitudes within the profile window to form the longitude 
resolved fluctuation spectra \citep[LRFS,][]{B73}. The LRFS for a sequence of 
pulses in the two emission modes of PSR B0844-35 is shown in 
Fig.~\ref{fig:lrfsB0844}. The periodic behaviour is seen as a peak frequency 
($f_p$) in the LRFS of mode A (left panel). The periodicity of subpulse 
drifting is $P_3$ (1/$f_p$) = 2.03$\pm$0.03$P$, which is visible as a odd-even 
drift in the single pulse sequence and is consistent with the earlier estimates
\citep{BMM16}. On the contrary no clear periodic behaviour is seen in the LRFS 
of mode B (Fig.~\ref{fig:lrfsB0844}, right panel), although there is a 
possibility of a wide feature between 0.3-0.4 $cy/P$, particularly in the 
leading side of the window. The presence of clear drifting behaviour cannot be 
confirmed in this mode. 

\begin{figure}
\plottwo{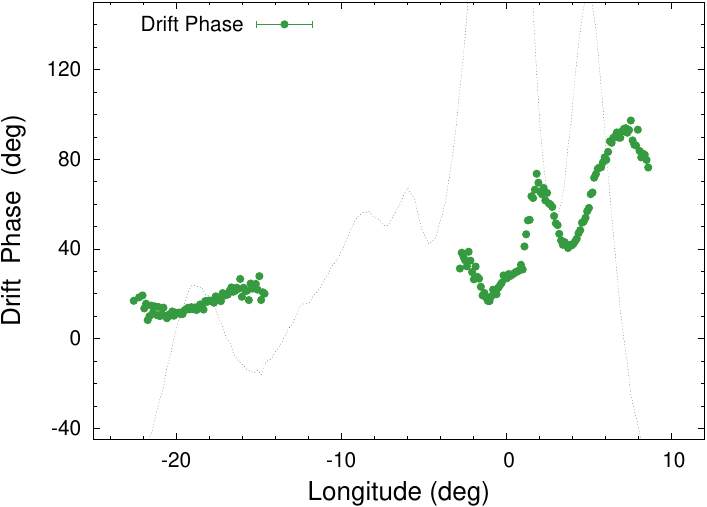}{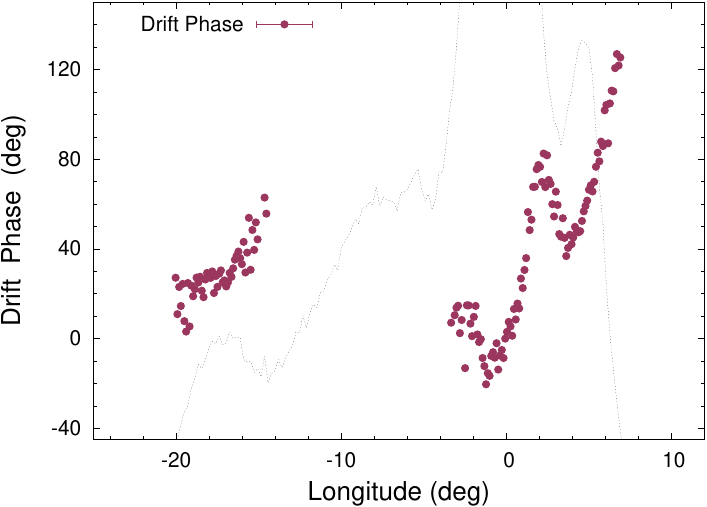}
\caption{The phase variations across the emission window corresponding to the
subpulse drifting behaviour in PSR B0844-35 during emission mode A. The left 
panel shows the phase variations in the low frequency range between 300 and 500
MHz while the high frequency behaviour between 550 and 750 MHz is shown in the 
right panel. The average profiles are shown in the background to highlight the 
phase behaviour across each component.
\label{fig:DriftphsB0844}}
\end{figure}

\begin{deluxetable}{cccccccccc}
\tablecaption{Subpulse Drifting Estimates in Mode A of PSR B0844-35 
\label{tab:driftB0844}}
\tablewidth{0pt}
\tablehead{
 \multicolumn{4}{c}{} & \multicolumn{2}{c}{\underline{Comp-1}} & \multicolumn{2}{c}{\underline{Comp-3}} & \multicolumn{2}{c}{\underline{Comp-4}} \\
 \colhead{Frequency} & \colhead{$f_p$} & \colhead{$FWHM$} & \colhead{$P_3$} & \colhead{$\Delta\phi$} & \colhead{$d\psi/d\phi$} & \colhead{$\Delta\phi$} & \colhead{$d\psi/d\phi$} & \colhead{$\Delta\phi$} & \colhead{$d\psi/d\phi$} \\
 \colhead{(MHz)} & \colhead{($cy/P$)} & \colhead{($cy/P$)} & \colhead{($P$)} & \colhead{(\degr)} & \colhead{(\degr/\degr)} & \colhead{(\degr)} & \colhead{(\degr/\degr)} & \colhead{(\degr)} & \colhead{(\degr/\degr)}}
\startdata
 300-500 & 0.492$\pm$0.007 & 0.017 & 2.03$\pm$0.03 & -20.7\degr : -15.0\degr & 2.8$\pm$0.1 & -1.1\degr : 1.0\degr & 7.2$\pm$0.5 & 3.8\degr : 7.1\degr & 18.0$\pm$0.6 \\
   &   &   &   &   &   &   &   &   &   \\
 550-750 & 0.492$\pm$0.006 & 0.015 & 2.03$\pm$0.03 & -20.0\degr : -15.1\degr & 5.7$\pm$0.6 & -1.4\degr : 1.0\degr & 15.4$\pm$1.5 & 3.6\degr : 6.9\degr & 27.8$\pm$1.1 \\
\enddata
\end{deluxetable}

The LRFS also measures the relative phase variations ($\psi$) of the subpulses 
across the longitude range of the profile (see top window in 
Fig.~\ref{fig:lrfsB0844}), that represents the relative motion of the sparks in
the IAR along the LOS cuts. The phase variations can be measured with more 
sensitivity using the time varying fluctuation spectra over the entire 
observing sequence \citep{BM18a}. The LRFS was initially estimated for 256 
pulses from the start of the observing session. The starting point was shifted
by 50 periods and the next set of 256 pulses were used for the LRFS, with the 
process continued till the end of the observing run. The average drift phase, 
at each pulse longitude, from all these independent measurements has been 
estimated as shown in Fig.~\ref{fig:DriftphsB0844}, for the two wide frequency 
bands 300-500 MHz (left panel) and 550-750 MHz (right panel). The pulsar 
profile at both frequency bands were aligned such that the central longitude 
($\phi=0\degr$) coincided with the profile peak, i.e. the peak of the trailing 
inner conal component. We have only included the average phase measurements in 
the longitudes where significant detection of the drifting behaviour were 
recorded. Hence, the phase measurements are missing in the central composite 
component consisting of the leading inner cone and the core, where drifting 
features are not clearly visible. The increased sensitivity of the wideband 
observations has revealed more details in the drifting phase behaviour, 
particularly at the boundary between the components, compared to previous 
studies \citep{BMM19}. 

he phase variations across the leading outer conal component show a smooth, 
shallow positive slope. The phase behaviour become more complicated in the 
trailing half of the profile comprising of the trailing inner and outer cones. 
On closer inspection two short sections can be seen where the phase variations 
are roughly linear that coincide with the location of the two component peaks. 
In case of the third component the linear sections are between $\phi=-1.1\degr$
and $\phi=1.0\degr$ at 300-500 MHz, and between $\phi=-1.4\degr$ and 
$\phi=1.0\degr$ at the 550-750 MHz band. While in the trailing fourth component
the linear behaviour is seen between $\phi=3.8\degr$ and $\phi=7.1\degr$ at the
300-500 MHz frequency band, and between $\phi=3.6\degr$ and $\phi=6.9\degr$ in 
the upper 550-750 MHz band profile. However, at the boundary of the different 
components the phase behaviour shows deviation from the linear behaviour. At 
the border between core and the inner cone, around $\phi=-2\degr$, and at 
trailing edge of the outer cone only for the lower frequency band, around 
$\phi=8\degr$, the phases show a turnover with a negative slope, while at the 
boundary between the inner cone and the outer cone, around $\phi = 2\degr$, the
phases show a bell shaped curve. 

The variations seen around the boundary do not reflect any systematic 
variations due to drifting, but arises due to the overlapping nature of these 
regions with contributions from two disjointed sides. In some single pulses, 
emission from one side dominate while at other times the other side becomes 
prominent, resulting in the bell shaped behaviour. Such phase behaviour with a 
bell shaped or U-shaped curve is also evident in some pulsars with periodic 
amplitude modulation, like PSR B0823+26, B1642-03, etc., where instead of 
showing systematic shift of the subpulses across the emission window, the 
entire emission becomes narrower and wider in a periodic manner \citep{BM19}.
Table \ref{tab:driftB0844} reports the measurement of the average drifting 
properties in the two frequency bands, including $f_p$, $P_3$ and the width of
the drifting frequency feature at 50 percent level of the peak value ($FWHM$). 
The table also lists the longitude range ($\Delta\phi$) across each profile
component where the phase behaviour shows linear nature, and the gradient of 
the phase variations across this longitude range ($d\psi/d\phi$). The phase 
gradients show evolution between the two frequency bands, with the high 
frequency measurements showing steeper gradients, almost a factor of two 
higher. The change in the drift phase behaviour with frequency has been 
reported in PSR B0809+74 \citep{HSW13}, having S$_d$ profile shape with 
peripheral LOS cuts across the emission beam. Our results show that such 
variations can also be detected in pulsars with central LOS traverse.

\begin{figure}
\plottwo{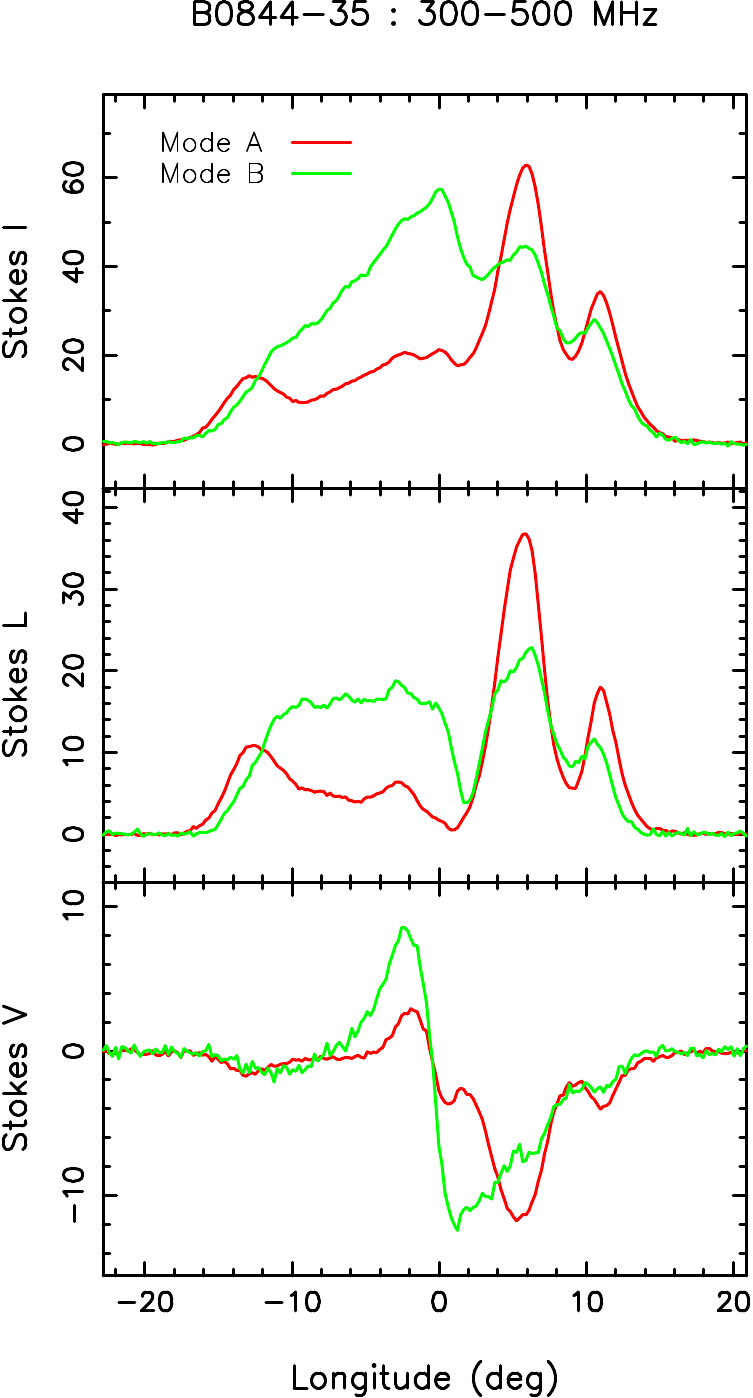}{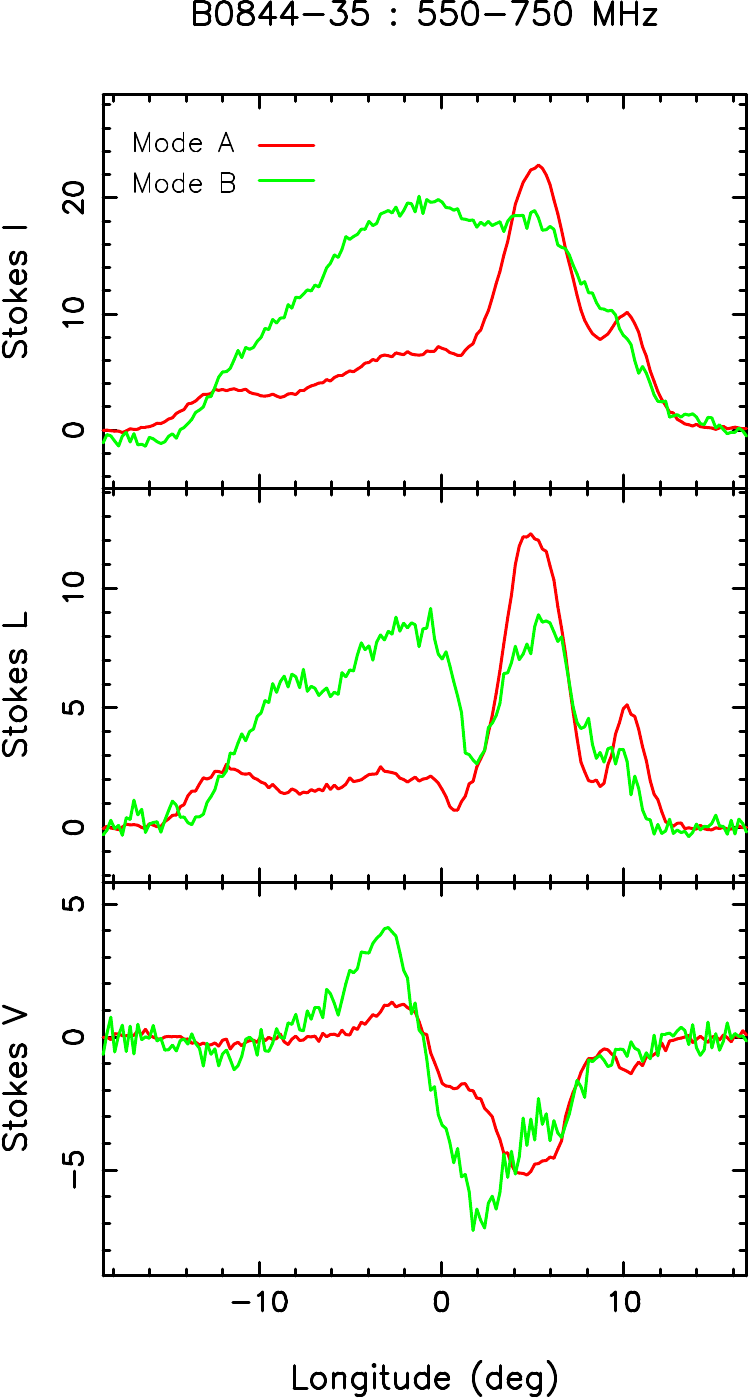}
\caption{The polarization behaviour across the average profiles of PSR 
B0844-35, with the two modes A and B shown in the same plot. The left panels
show the average profiles over the 300-500 MHz frequency range and the right 
panels show the corresponding profiles averaged between 550-750 MHz. The top
window in each panel represents the total intensity profiles, the middle window
shows the variation of the linear polarization across the profile and the 
bottom window the circular polarization behaviour.
\label{fig:polB0844}}
\end{figure}

\begin{figure}
\gridline{\fig{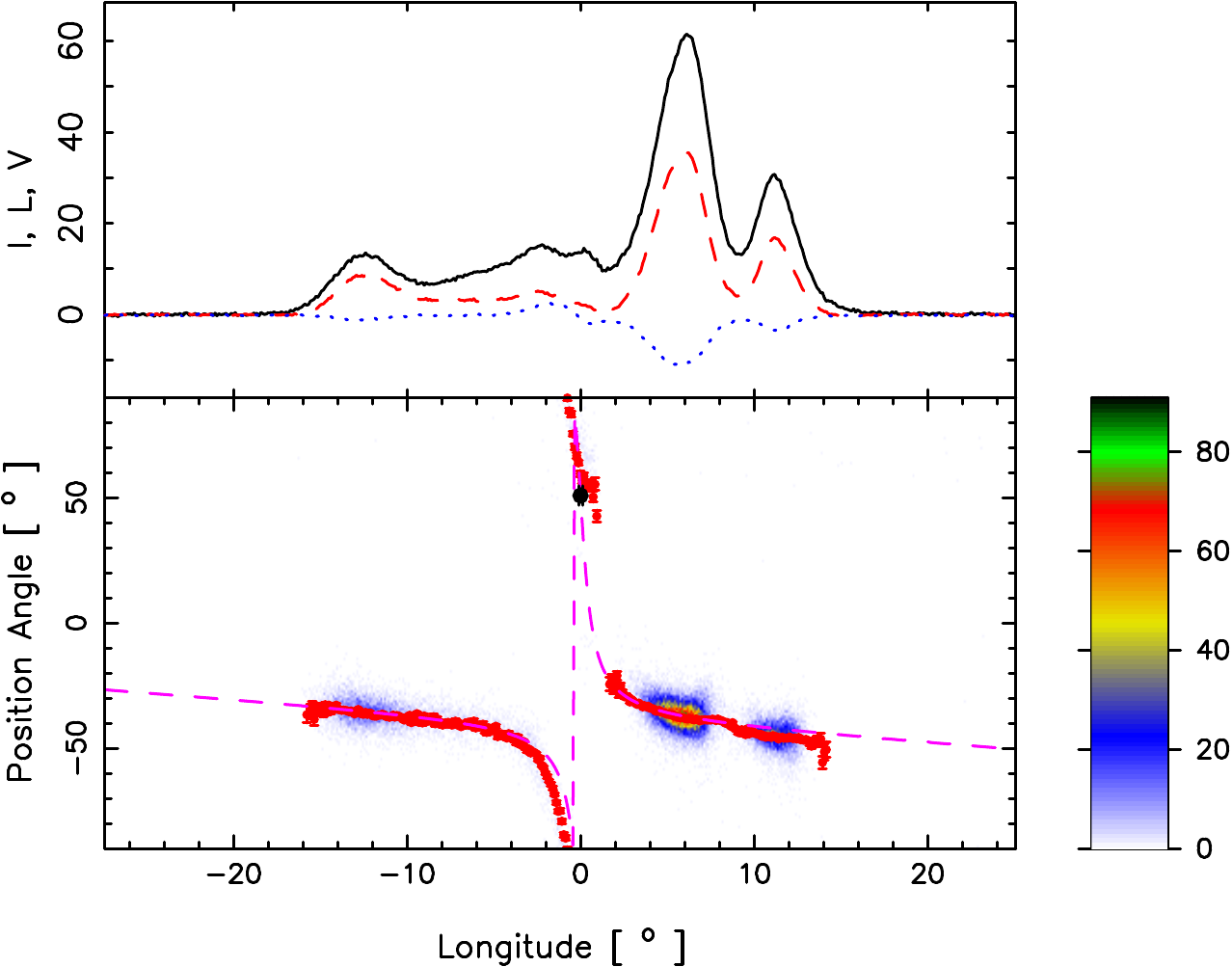}{0.45\textwidth}{(a) Mode A, 300-500 MHz}
          \fig{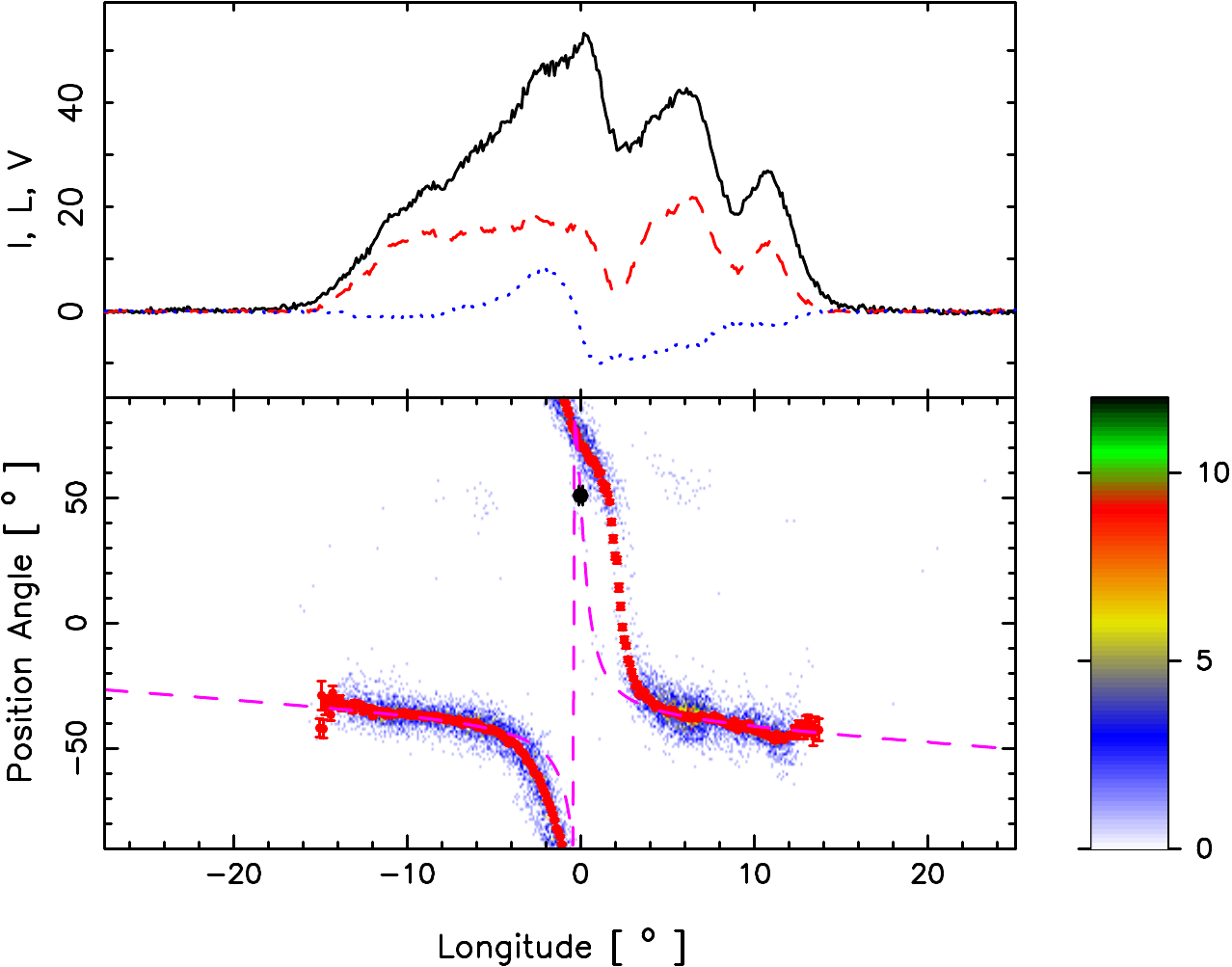}{0.45\textwidth}{(b) Mode B, 300-500 MHz}
          }
\gridline{\fig{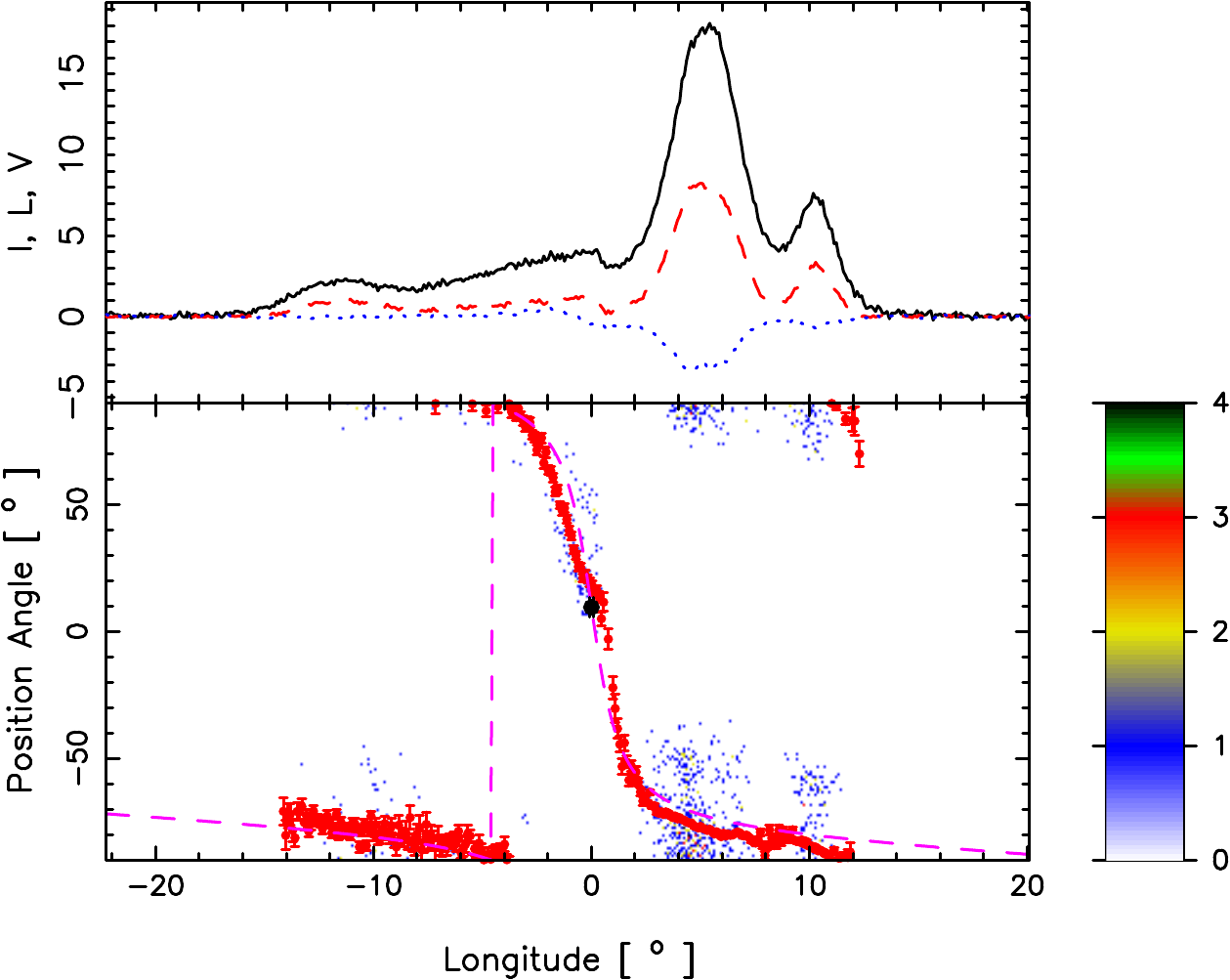}{0.45\textwidth}{(c) Mode A, 550-750 MHz}
          \fig{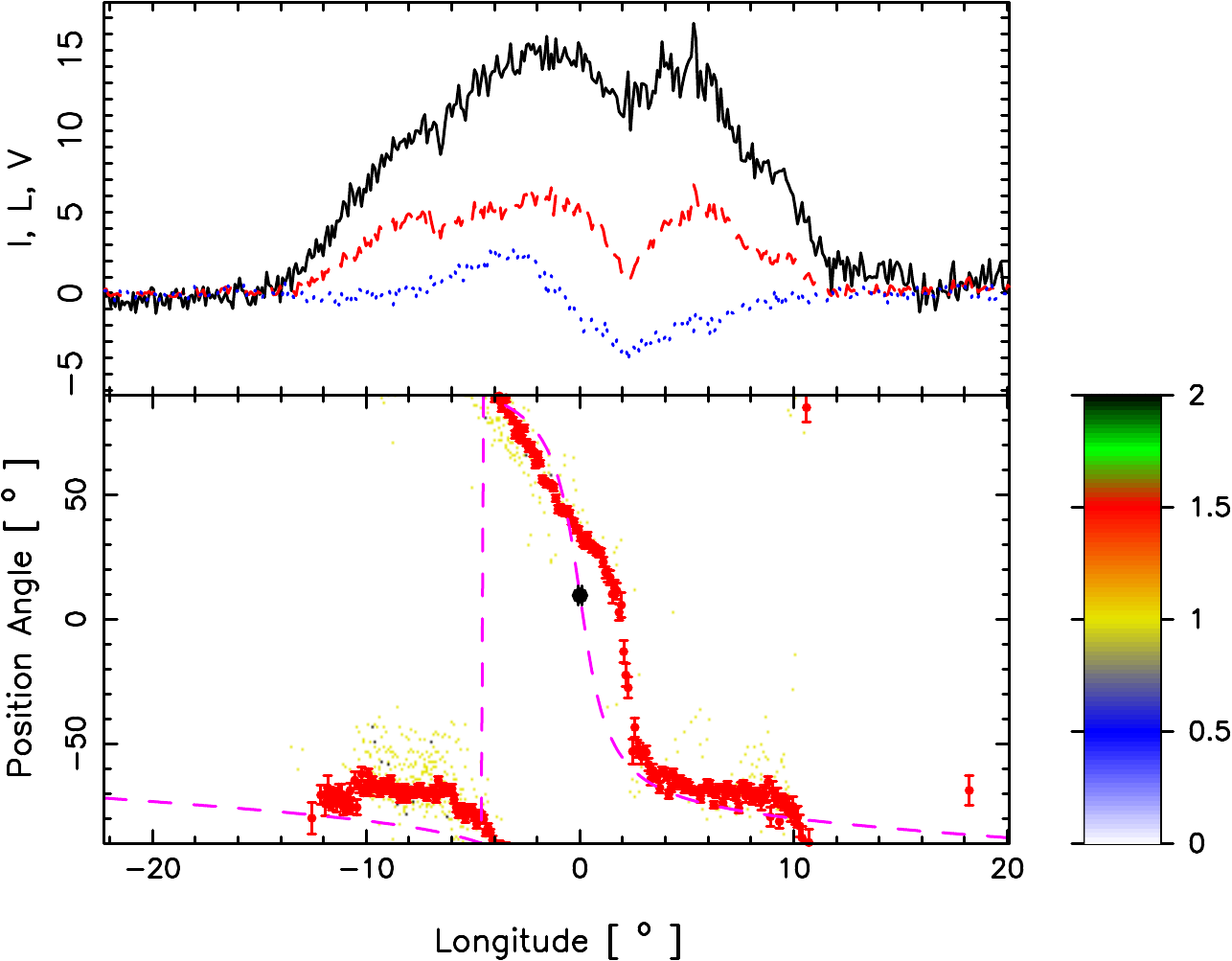}{0.45\textwidth}{(d) Mode B, 550-750 MHz}
          }
\caption{The figure shows the polarization position angle (PPA) distribution of
the single pulse time samples of PSR B0844-35. The top windows (a) and (b) 
shows the behaviour of mode A and B, respectively, at the 300-500 MHz frequency
range. The bottom windows (c) and (d) shows the PPA of the two modes averaged 
between 550 and 750 MHz. The top panel in each window shows the average 
profile (black), the average linear (red) and circular polarization (blue) 
properties of the radio emission. In the bottom window the distribution of the 
single pulse PPA is shown as a colour scale, representing the number of points 
in each location, as well as the average PPA behaviour across the window (red 
error bars). The rotating vector model (RVM) fits to the PPA is also shown in 
the figure (pink dashed line), along with the steepest gradient point of the 
RVM fit (black error bar).
\label{fig:B0844ppa}}
\end{figure}

\subsection{Polarization Properties and Emission Height}
The polarization behaviour across the average profiles of PSR B0844-35 is shown
in Fig.~\ref{fig:polB0844}, where both emission modes are shown in the same
plot, with left panel showing the average behaviour over 300 to 500 MHz 
frequency range while the right panel between 550 and 750 MHz. Each panel in 
the figure comprises of three parts, representing the total intensity profiles 
(top window), the linear polarization (middle window), and the circular 
polarization (bottom window) across the two profiles. The presence of the core
emission at the center of the profile is highlighted by the decrease in the
linear polarization level as well as change in the sign of the circular 
polarization \citep{MRG07,SRM13}. The profiles of both emission modes at each 
frequency range have been aligned with respect to a common center, identified 
from the polarization position angle (see below) to compare their relative 
locations. The figure shows that the emission window, including the location of
the individual components, remains the same in both modes A and B, suggesting 
that the location of the radio emission is not affected by mode changing in 
this pulsar. 

\begin{deluxetable}{ccccccccc}
\tablecaption{Estimation of radio emission height in PSR B0844-35
\label{tab:emhtB0844}}
\tablewidth{0pt}
\tablehead{
 \colhead{Mode} & \colhead{Frequency} & \colhead{SG} & \colhead{$\phi_{\circ}$} & \colhead{$\phi_l$} & \colhead{$\phi_t$} & \colhead{$\phi_c$} & \colhead{$\Delta\phi$} & \colhead{$h_{\rm A/R}$} \\
    & \colhead{(MHz)} & \colhead{(\degr/\degr)} & \colhead{(\degr)} & \colhead{(\degr)} & \colhead{(\degr)} & \colhead{(\degr)} & \colhead{(\degr)} & \colhead{(km)}}
\startdata
  A & 300-500 & -90.0 & 0.14$\pm$0.06 & -18.10$\pm$0.21 & 17.00$\pm$0.21 & -0.55$\pm$0.30 & 0.69$\pm$0.31 & 160$\pm$72 \\
    & 550-750 & -60.0 & 0.04$\pm$0.14 & -16.44$\pm$0.21 & 13.58$\pm$0.21 & -1.43$\pm$0.30 & 1.47$\pm$0.33 & 342$\pm$77 \\
    &   &   &   &   &   &   &   &   \\
  B & 300-500 & -112.0 & 0.75$\pm$0.06 & -17.04$\pm$0.21 & 15.94$\pm$0.21 & -0.55$\pm$0.30 & 1.30$\pm$0.31 & 302$\pm$72 \\
    & 550-750 & -39.9 & 1.21$\pm$0.12 & -13.87$\pm$0.21 & 13.40$\pm$0.21 & -0.24$\pm$0.30 & 1.45$\pm$0.32 & 337$\pm$74 \\
\enddata
\end{deluxetable}

The polarization position angle (PPA) for the single pulse time samples with 
significant polarization detection is shown in Fig.~\ref{fig:B0844ppa} 
\citep[see][for more details about making these plots]{MMB23a,MMB23b}. The 
average PPA resemble the characteristic S-shaped curve that follows the 
rotating vector model \citep[][]{RC69}, although certain deviation from the RVM
nature is seen particularly in the core region due to presence of orthogonal
polarization modes \citep{MRG07,SRM13,BMR19,MMB23a,MMB23b}. The single pulse 
plots of the PPA generally show time samples with significant detection of 
polarization power. In case not enough such time samples are available, the 
orthogonal polarization modes are not clearly visible in these plots, although
the deviations of the average profile PPAs from the RVM can still be a result 
of these invisible orthogonal mode mixing. More sensitive observations are 
required in such cases to detect the polarization modes in the single pulse 
plots. 

The RVM, in the static dipole approximation, gives an estimate of the PPA, 
$\chi$, as a function of the pulse longitude, $\phi$, using the angle between 
the magnetic and rotation axes, $\alpha$, and the angle of closest approach of 
the LOS to the magnetic axis, 
$\beta$, in the form :
\begin{equation}
\chi = \chi_{\circ} + \tan^{-1} \left( \frac{\sin{\alpha}~\sin{(\phi-\phi_{\circ}})} {\sin{(\alpha+\beta)}\cos{\alpha} - \sin{\alpha}\cos{(\alpha+\beta)}\cos{(\phi-\phi_{\circ})}}\right)
\label{eq:RVM}
\end{equation}
Here, $\chi_{\circ}$ and $\phi_{\circ}$ are the arbitrary phase offsets in the
PPA and the profile longitude, respectively. The RVM model for the PPA are also
shown in Fig.~\ref{fig:B0844ppa} (pink dashed line) and closely matches the 
observed behaviour, except the core region in mode B. Although the geometrical
angles, $\alpha$ and $\beta$, cannot be fixed from RVM fits due to highly 
correlated solutions \citep{EW01,ML04}, the steepest gradient (SG) of the RVM,
$\mid d\chi/d\phi \mid_{max}$, and its location specified by $\phi_{\circ}$ 
(black point with error bar in Fig.~\ref{fig:B0844ppa}) can be used to find 
estimates of the radio emission height. It has been shown that for the emission
heights less than 10\% of the light cylinder distance, the aberration and 
retardation effects, arising from pulsar rotation, causes a positive shift in 
$\phi_{\circ}$ with respect to the the center of the profile, $\phi_c$, and the
shift $\Delta \phi = \phi_{\circ}-\phi_c$, is proportional to the emission 
height, $h_{A/R}$, \citep{BCW91,D08}
\begin{equation}
h_{\rm A/R} = \frac{c P \Delta\phi}{8\pi} \approx 208.2~\text{km} \left(\frac{P}{\text{s}}\right) \left(\frac{\Delta\phi}{\text{deg}}\right)
\label{eq:hAR}
\end{equation}
The estimation of $\phi_c$ requires identifying two longitudes, $\phi_l$ and 
$\phi_t$, at the leading and trailing edge of the profile corresponding to the
last open field line, such that $\phi_c = \phi_l + (\phi_t - \phi_l)/2$. In 
practice $\phi_l$ and $\phi_t$ are identified as the longitudes on either edge
of the profile that are above the detection limit, i.e. above five times the
baseline noise rms level. Table \ref{tab:emhtB0844} shows the measurement of 
the relevant quantities and emission heights for the two modes A and B, 
estimated over the two observing frequency ranges, 300-500 MHz and 550-750 MHz.
In both modes the radio emission originates a few hundred kilometers above the
neutron star surface, that is consistent with the height estimates in the 
normal pulsar population \citep{M17,MMB23b}. The height estimates are affected 
by errors arising from identifying the SG point due to depolarization of the 
core component, particularly in mode B, as well as the identification of the
profile edges due to detection sensitivities. Although the wideband 
measurements provide improved sensitivity for the edge detections, the emission
heights should be used as suggestive for the location of the radio emission in 
the pulsar magnetosphere, rather than exact estimates.

\begin{figure}
\epsscale{0.75}
\plottwo{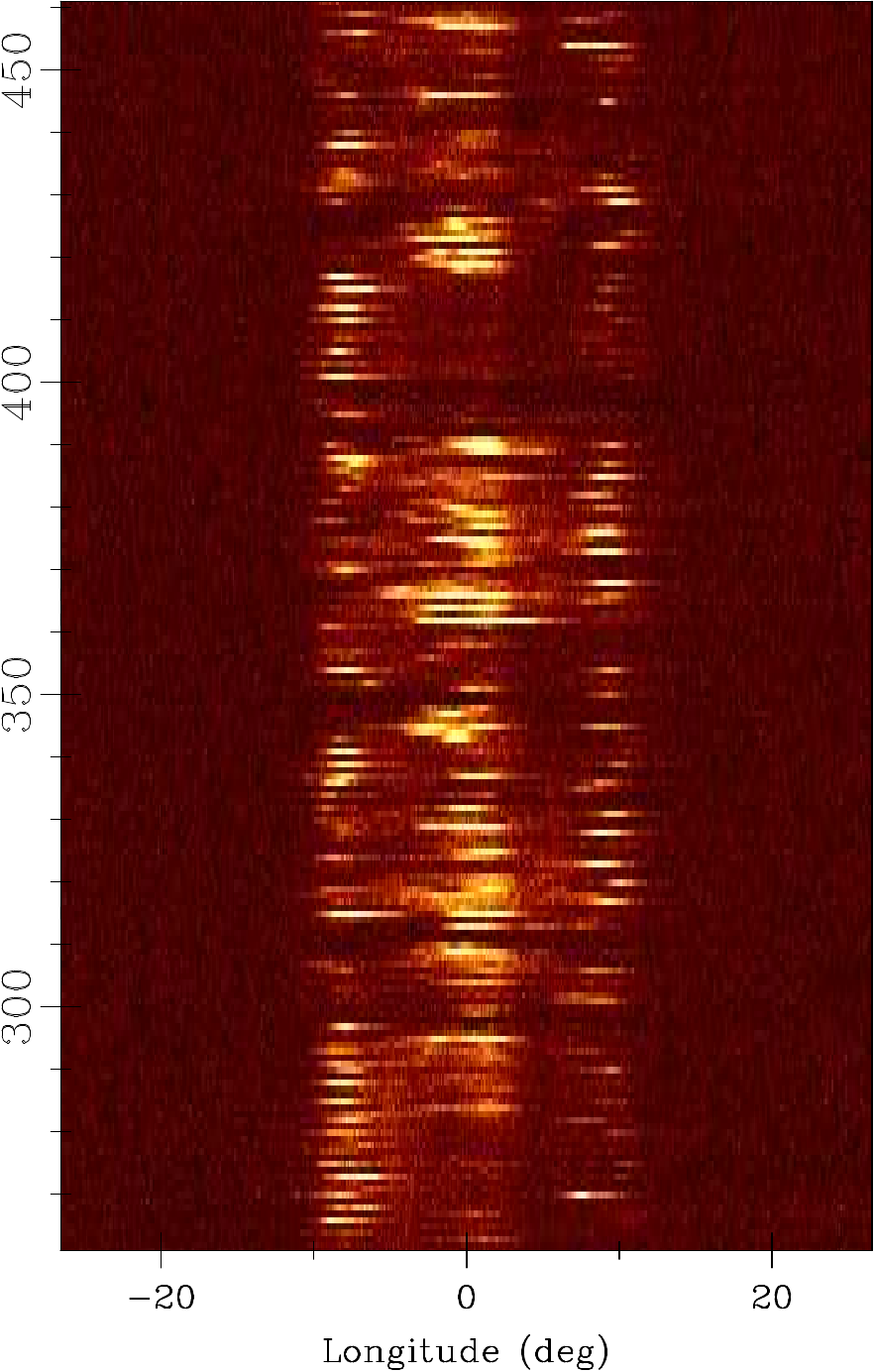}{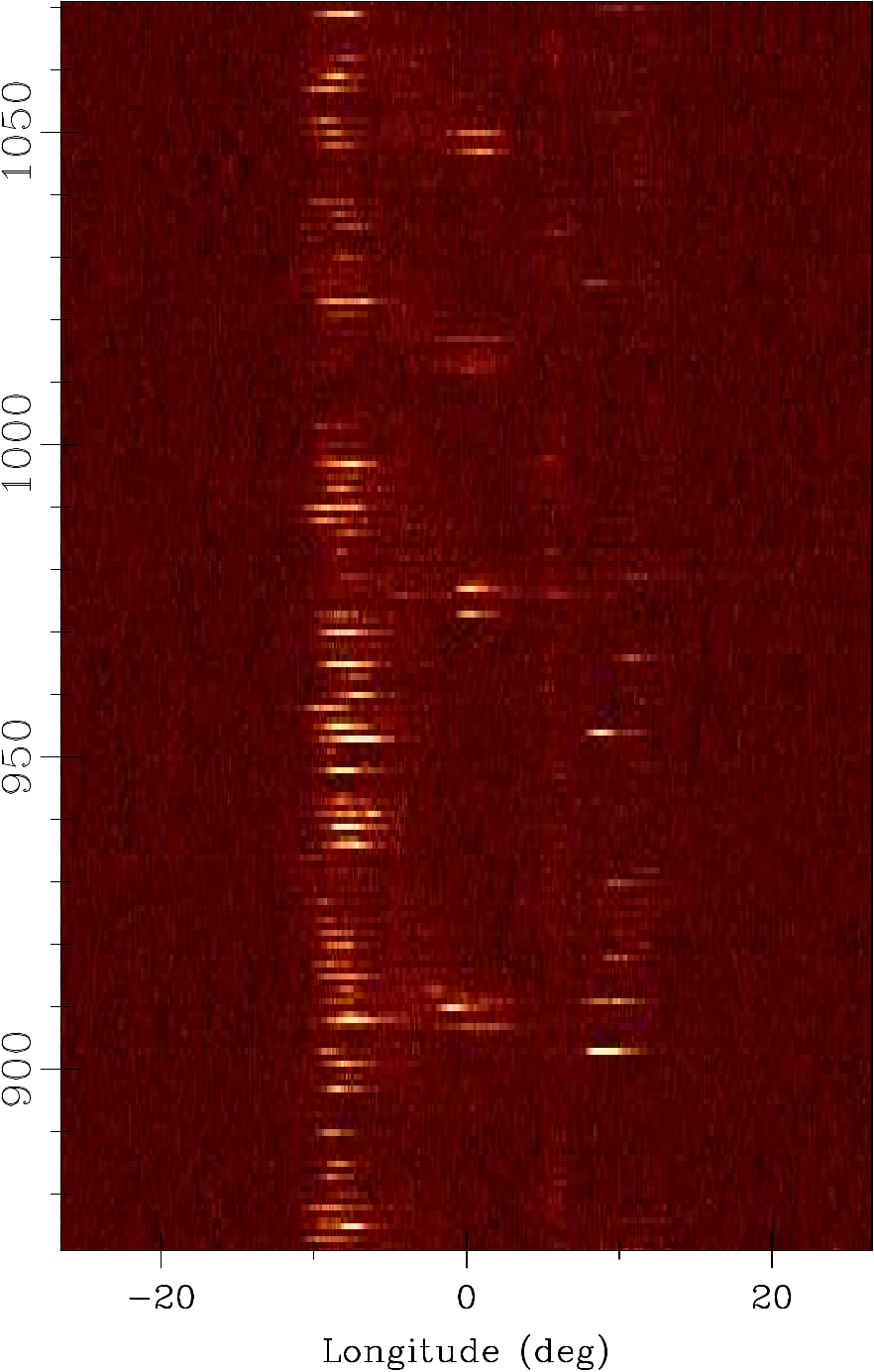}
\caption{The left panel shows the single pulse sequence corresponding to the 
Bright mode of PSR B1758-29, observed on 18 March, 2019, and represents the 
pulse range between 260 to 460 from the start of the observing session. The 
right panel shows the pulse sequence between pulses 870 and 1070, belonging to 
the Quiet mode.
\label{fig:singlB1758}}
\end{figure}

\begin{figure}
\plottwo{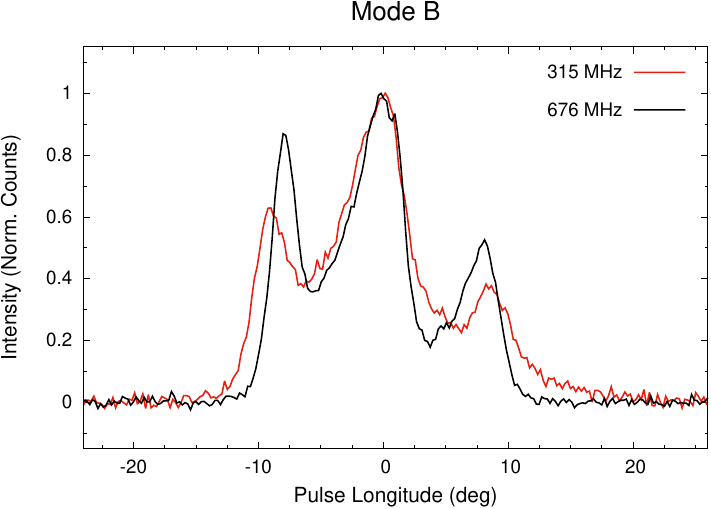}{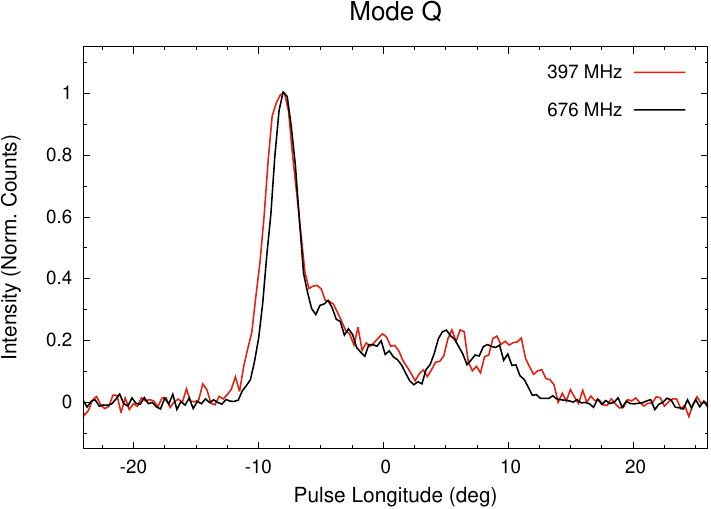}
\caption{The average profiles of PSR B1758-29 during B mode (left panel) and
Q mode (right panel). The profiles at two separated frequencies are shown in
each panel to highlight the frequency evolution of the modes.
\label{fig:profB1758}}
\end{figure}

\begin{deluxetable}{ccccccccccc}
\tablecaption{Mode Sequence of PSR B1758-29 \label{tab:seqB1728}}
\tablewidth{0pt}
\tablehead{
 \multicolumn{3}{c}{\underline{18/03/2019}} &   &   & \multicolumn{6}{c}{\underline{10/02/2020}} \\
 \colhead{Pulse Range} & \colhead{Mode} & \colhead{Mode Length} &   &   & \colhead{Pulse Range} & \colhead{Mode} & \colhead{Mode Length}  & \colhead{Pulse Range} & \colhead{Mode} & \colhead{Mode Length} \\
 \colhead{($P$)} &   & \colhead{($P$)} &   &   & \colhead{($P$)} &   & \colhead{($P$)} & \colhead{($P$)} &   & \colhead{($P$)}}
\startdata
    1 - 185  & Q & 185 &   &   &    1 - 298  & B & 298 & 3105 - 3228 & B & 124 \\
  186 - 650  & B & 465 &   &   &  299 - 500  & Q & 202 & 3229 - 3302 & Q &  74 \\
  651 - 720  & Q &  70 &   &   &  501 - 955  & B & 455 & 3303 - 3336 & B &  34 \\
  721 - 800  & B &  80 &   &   &  956 - 1097 & Q & 142 & 3337 - 3936 & Phasing & 600 \\
  801 - 1181 & Q & 381 &   &   & 1098 - 1259 & B & 162 & 3937 - 4103 & B & 167 \\
 1182 - 1531 & B & 350 &   &   & 1260 - 1351 & Q &  92 & 4104 - 4256 & Q & 153 \\
 1532 - 1608 & Q &  77 &   &   & 1352 - 1845 & B & 494 & 4257 - 4653 & B & 397 \\
 1609 - 2100 & B & 492 &   &   & 1846 - 2092 & Q & 247 & 4654 - 4769 & Q & 117 \\
   &  &  &   &   & 2093 - 2185 & B &  93 & 4770 - 4863 & B &  94 \\
   &  &  &   &   & 2186 - 2297 & Q & 112 & 4864 - 4954 & Q &  91 \\
   &  &  &   &   & 2298 - 2331 & B &  35 & 4955 - 5064 & B & 110 \\
   &  &  &   &   & 2332 - 2455 & Q & 124 & 5065 - 5141 & Q &  77 \\
   &  &  &   &   & 2456 - 2696 & B & 241 & 5142 - 5240 & B &  99 \\
   &  &  &   &   & 2697 - 2930 & Q & 234 & 5241 - 5694 & Q & 454 \\
   &  &  &   &   & 2931 - 3002 & B &  72 & 5695 - 5871 & B & 177 \\
   &  &  &   &   & 3003 - 3104 & Q & 102 \\
\enddata
\end{deluxetable}

\begin{figure}
\epsscale{0.6}
\plotone{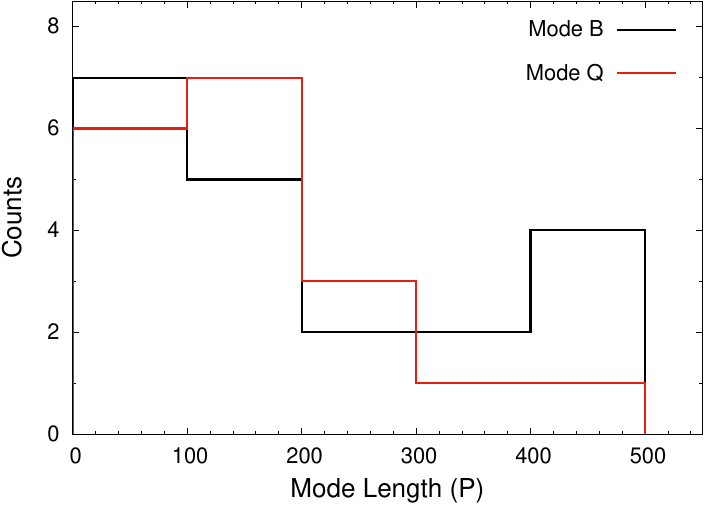}
\caption{The distribution of the durations of the two emission modes of PSR 
B1758-29.
\label{fig:lendistB1758}}
\end{figure}

\begin{deluxetable}{ccccccc}
\centerwidetable
\tablecaption{Average Profile properties in the emission modes of PSR B1758-29 \label{tab:profB1758}}
\tablewidth{0pt}
\tablehead{
 \colhead{Mode} & \colhead{Frequency} & \colhead{$W_{5\sigma}$} & \colhead{$W_{10}$} & \colhead{$W_{50}$} & \colhead{$W_{sep}^{in}$} & \colhead{$W_{sep}^{out}$} \\
    & \colhead{(MHz)} & \colhead{(\degr)} & \colhead{(\degr)} & \colhead{(\degr)} & \colhead{(\degr)} & \colhead{(\degr)}}
\startdata
     & 315 & 28.8$\pm$0.6 & --- & 21.3$\pm$0.5 & --- & 17.5$\pm$0.5 \\
     & 345 & 25.9$\pm$0.6 & --- & 20.7$\pm$0.5 & --- & 17.4$\pm$0.5 \\
     & 397 & 25.5$\pm$0.6 & --- & 20.1$\pm$0.5 & --- & 17.1$\pm$0.5 \\
     & 433 & 25.7$\pm$0.6 & 23.0$\pm$0.5 & 19.6$\pm$0.5 & --- & 16.9$\pm$0.5 \\
  B  & 468 & 24.6$\pm$0.6 & 22.3$\pm$0.5 & 19.4$\pm$0.5 & --- & 16.7$\pm$0.5 \\
     & 576 & 22.9$\pm$0.6 & --- & 18.9$\pm$0.5 & --- & 16.2$\pm$0.5 \\
     & 610 & 23.1$\pm$0.6 & 21.3$\pm$0.5 & 18.8$\pm$0.5 & --- & 16.0$\pm$0.5 \\
     & 643 & 22.6$\pm$0.6 & 21.2$\pm$0.5 & 18.6$\pm$0.5 & --- & 16.0$\pm$0.5 \\
     & 676 & 22.9$\pm$0.6 & 21.2$\pm$0.5 & 18.6$\pm$0.5 & --- & 15.9$\pm$0.5 \\
     & 709 & 22.7$\pm$0.6 & 21.0$\pm$0.5 & 18.5$\pm$0.5 & --- & 15.8$\pm$0.5 \\
     &     &   &   &   &   &   \\
     & 315 & 24.0$\pm$0.6 & --- & --- & --- & --- \\
     & 345 & 25.3$\pm$0.6 & --- & --- & --- & --- \\
     & 397 & 25.1$\pm$0.6 & --- & --- & --- & 18.1$\pm$0.5 \\
     & 433 & 25.3$\pm$0.6 & --- & 21.6$\pm$0.5 & 10.4$\pm$0.5 & 17.6$\pm$0.5 \\
  Q  & 468 & 24.0$\pm$0.6 & --- & 21.3$\pm$0.5 & 10.3$\pm$0.5 & 17.7$\pm$0.5 \\
     & 576 & 23.3$\pm$0.6 & --- & 20.7$\pm$0.5 & 9.6$\pm$0.5 & 17.0$\pm$0.5 \\
     & 610 & 23.5$\pm$0.6 & --- & 20.4$\pm$0.5 & 9.7$\pm$0.5 & 17.0$\pm$0.5 \\
     & 643 & 23.3$\pm$0.6 & --- & 20.0$\pm$0.5 & 9.6$\pm$0.5 & 16.8$\pm$0.5 \\
     & 676 & 23.3$\pm$0.6 & --- & 20.2$\pm$0.5 & 9.7$\pm$0.5 & 16.6$\pm$0.5 \\
     & 709 & 23.1$\pm$0.6 & --- & 20.1$\pm$0.5 & 9.7$\pm$0.5 & 16.6$\pm$0.5 \\
\enddata
\end{deluxetable}

\section{Radio emission features of PSR B1758-29}\label{sec:B1758-29}

\subsection{Emission States}
PSR B1758-29 was observed at 325 MHz and 610 MHz as part of the MSPES sample
for around 2000 pulses at both frequencies \citep{MBM16}. The average profile 
was classified as type T with a central core component and one pair of conal 
emission on either side of the core. \citet{BMM21} reported the presence of two
emission modes in the MSPES single pulse sequence, classified as bright (B) 
mode and quiet (Q) modes. The emission in the central core region dominates 
during mode B but becomes greatly reduced during the Q mode. Further analysis 
of the mode behaviour, including mode statistics, was not possible due to the 
the weaker detection sensitivity of the single pulse emission, particularly 
during mode Q, from these narrow band observations. The higher sensitivity 
wideband observations have allowed a detailed study of mode changing in this 
pulsar. 

Fig.~\ref{fig:singlB1758} shows two pulse sequences representing the two 
emission states, mode B (left panel) with bright central core, and mode Q 
(right panel) where both the core and the trailing conal components become much
weaker, average intensity being less than 20\% of the the leading cone. The 
single pulse behaviour of the two modes are not uniform but show significant 
variations over the observing durations. Both the core and conal emission 
become weaker at certain intervals during mode B, not in a co-ordinated manner.
On the other hand there are short bursts lasting a few periods when the core 
and the trailing conal emission becomes brighter during mode Q, independent of 
each other. The average profile of the two modes and their variation with 
frequency is shown in Fig.~\ref{fig:profB1758}, and clearly shows the dominant 
core during mode B and the leading cone dominated profile in the Q mode. The Q 
mode figures also shows the presence of the weaker inner cones that are 
otherwise hidden by the dominant core in the B mode, and the pulsar shows both 
pairs of inner and outer cones across the emission window. The B mode profile 
also highlights the frequency evolution of the intensity of the core and the 
conal emission, where the cores have a steeper spectra compared to the cones, 
with a relative difference in the spectral index $\Delta\alpha_{\rm core/cone} 
= -0.9$ \citep[see][for details]{BMM21,BMM22a}.

After perusing through the single pulses on each observing session we have 
identified the sequences of the two modes as reported in Table 
\ref{fig:profB1758}, while Fig.~\ref{fig:lendistB1758} shows the mode length 
distributions, combining the two observing sessions. The average statistics of
the modes are reported in Table \ref{tab:obs} and unlike PSR B0844-35 the 
durations of the two modes are more comparable, with mode B being about 40-50 
percent longer than the Q mode on an average. PSR B1758-29 spends around 60 
percent of its time in mode B and the remaining 40 percent in mode Q. The mode 
lengths usually lasted for a few hundred pulses for both modes and did not 
exceed 500 pulses during the two observations (see Fig.~\ref{fig:lendistB1758}).

The evolution of the profile widths as a function of frequency during the two
emission states is reported in Table \ref{tab:profB1758}, including the width
estimated at 5 times the noise rms level ($W_{5\sigma}$), the width at 10
percent ($W_{10}$) and 50 percent ($W_{50}$) of the peak intensity of the outer
most components on either side of the profile, and the separation between the 
outer ($W_{sep}^{out}$) conal pair. The inner conal pair was visible in the Q 
mode and we estimated the separation between them ($W_{sep}^{in}$), however, 
due to the lower intensity of the trailing outer cone the $W_{10}$ could not be
estimated in this case. The effect of radius to frequency is once again visible
in the frequency evolution of $W_{10}$, $W_{50}$ and $W_{sep}^{out}$. The 
separation between the inner conal pair usually do not show any change with 
frequency. At the higher frequency range above 550 Hz $W_{sep}^{in}$ remains 
constant, but shows a jump below this frequency. This is due to the difficulty
in identifying the location of the inner conal component peaks as the they 
become relatively weaker, compared to the core and the outer cones at lower 
frequencies, and merge with the outer cones. The $W_{50}$ and $W_{sep}^{out}$ 
estimates show that the profile window becomes wider as the pulsar transitions 
from the B mode to the Q mode. 

\begin{figure}
\plottwo{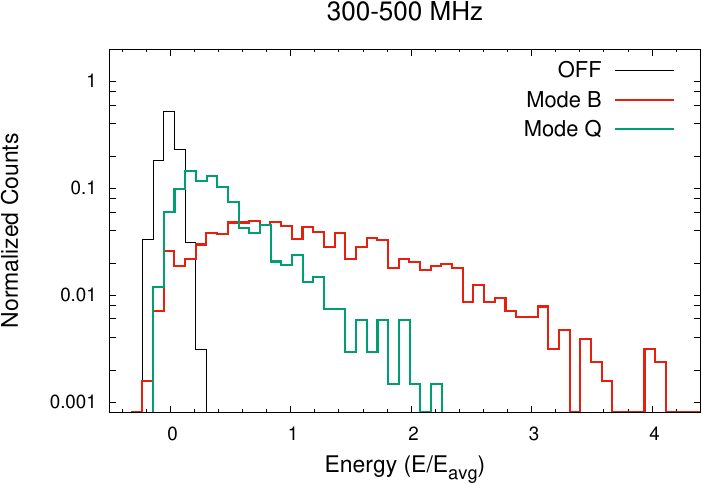}{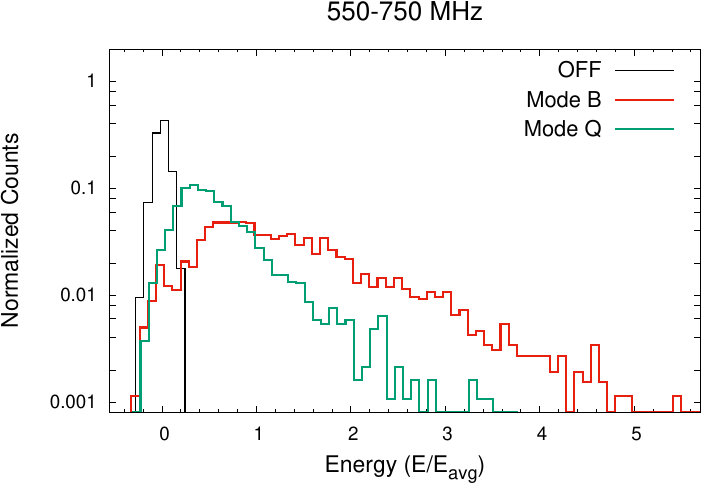}
\caption{The distribution of the average energies of single pulses in the two 
modes of PSR B1758-29. The left panel shows the distribution in the lower 
frequency band, between 300 and 500 MHz, while the upper band distribution, 
averaged over 550 and 750 MHz, is shown in the right panel. The Gaussian noise 
distribution in the off-pulse region is also shown in each plot.
\label{fig:distB1758}}
\end{figure}

The average energy distribution of the single pulses in the B and the Q modes
is estimated for the two frequency bands and shown in Fig.~\ref{fig:distB1758}.
The pulse energies show a log-normal distribution in both emission states with
a prominent tail. The plots also highlight the B mode to be more energetic than
the Q mode with a much longer tail in the distribution. The distribution of the
B mode at both frequencies show a bimodal structure with a second peak 
coincident with the baseline noise distribution, estimated in the off-pulse
region, suggesting the presence of nulling. However, the presence of numerous
lower energy pulses did not allow the identification of the individual null
pulses within the B mode sequences, either using statistical estimates or 
visual inspection. The nulling fractions can be estimated using Gaussian 
fitting techniques to the off-pulse and the null distributions of the B mode 
\citep{BMM17}. We found the upper limits of the nulling fraction to be 
4.9$\pm$0.6 percent at the lower frequency and 4.1$\pm$0.4 percent at the 
higher frequency. The nulling fractions over the two observing sessions are 
consistent within measurement errors and the variations can be attributed to 
the different observing durations and detection sensitivities. There were no 
clear separation between the null and the burst pulses in the Q mode 
distributions. Although it is likely that the Q mode also exhibits nulling, 
more sensitive observations are required to study the nulling behaviour in this
mode.

\begin{figure}
\gridline{\fig{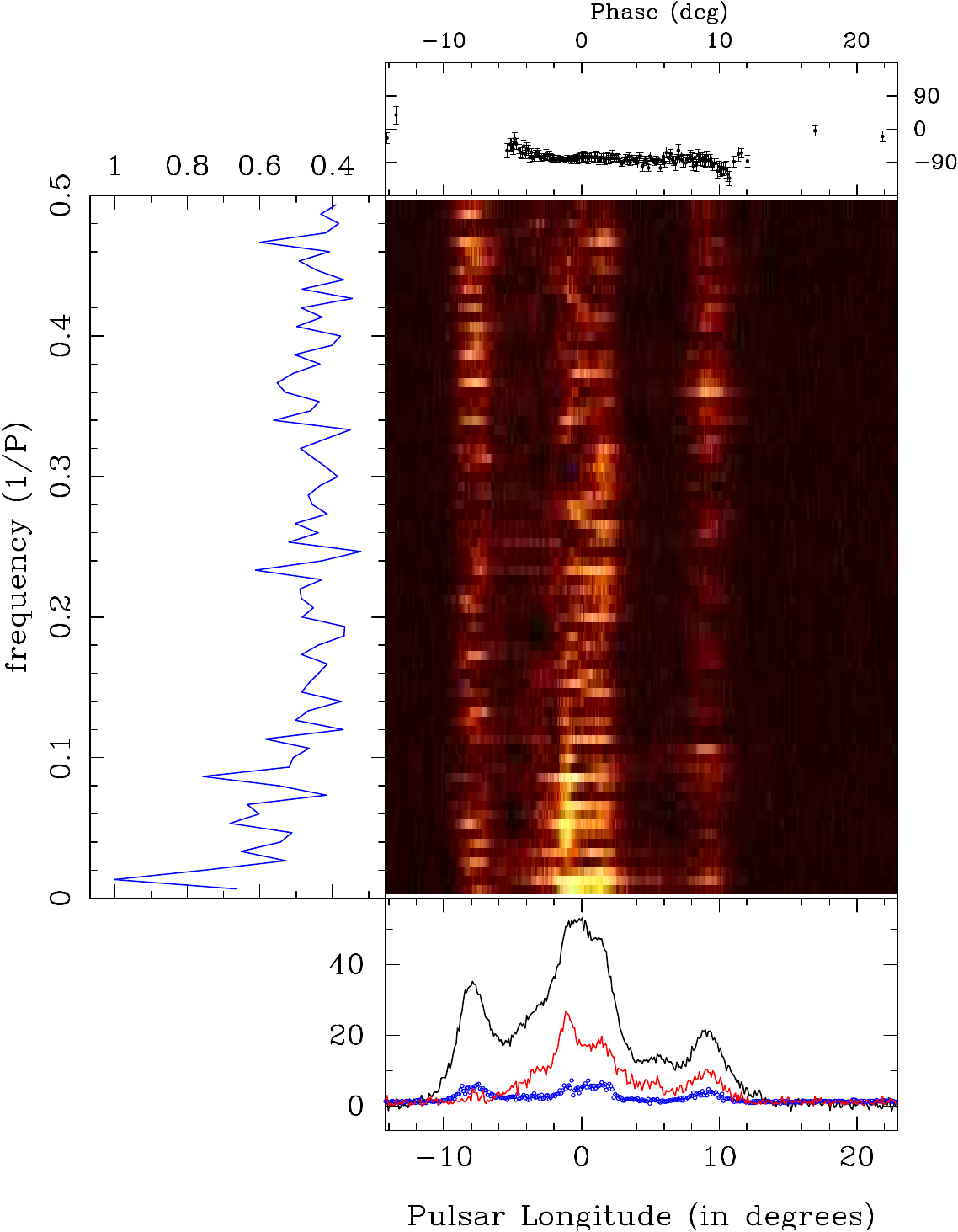}{0.41\textwidth}{(a)}
          \fig{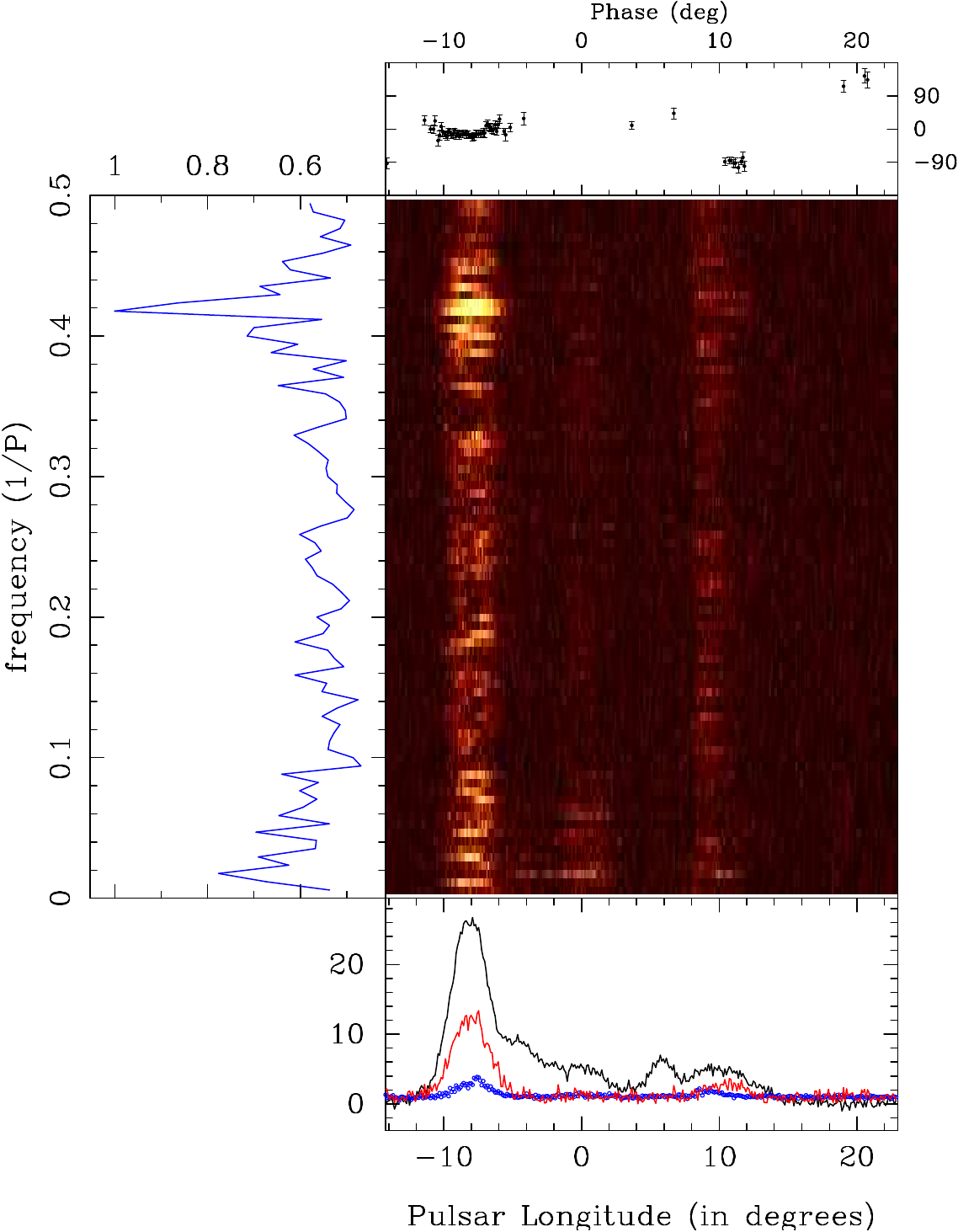}{0.41\textwidth}{(b)}
          }
\gridline{\fig{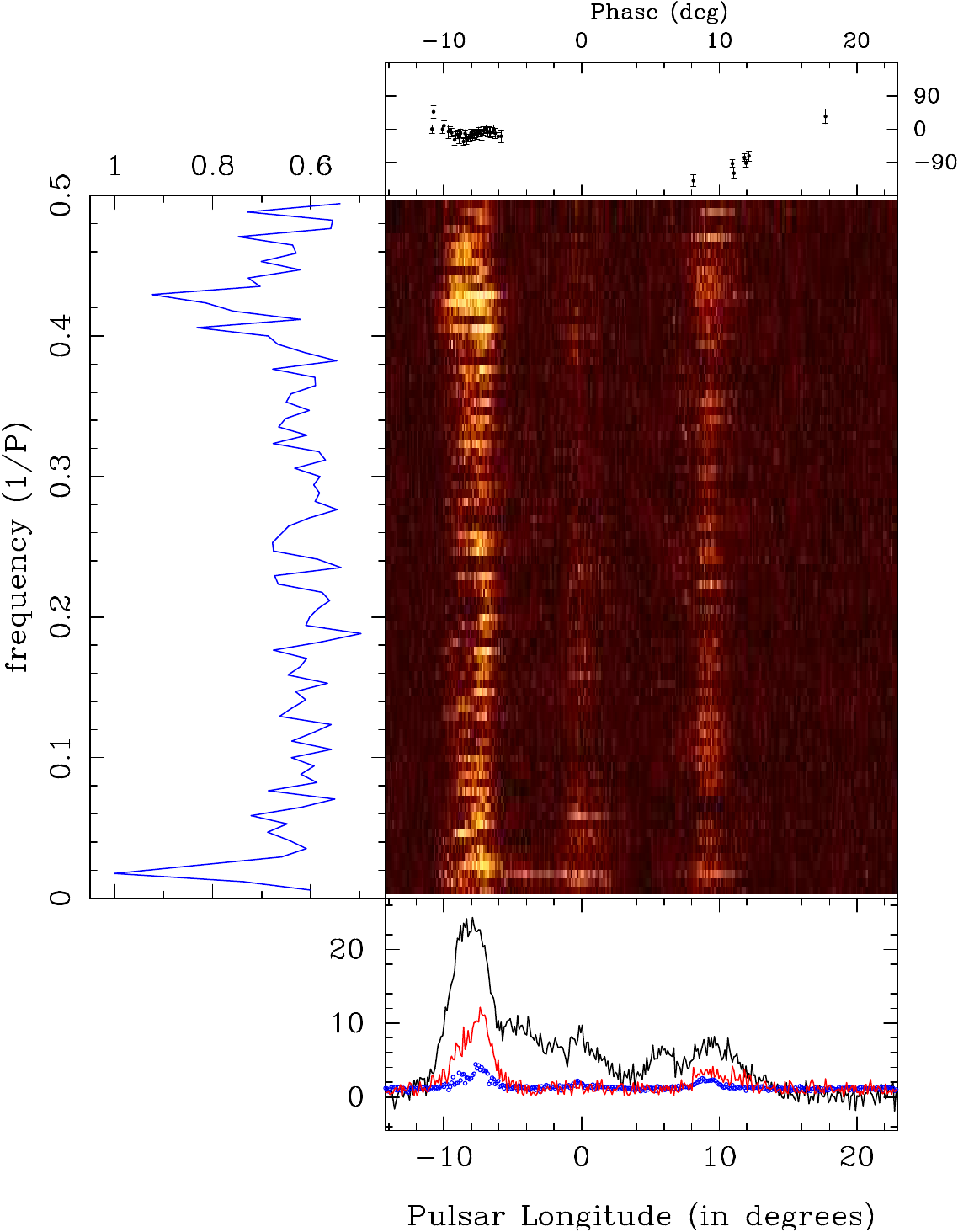}{0.41\textwidth}{(c)}
          }
\caption{The LRFS across the single pulse sequences of PSR B1758-29 during
the observations on 18 March, 2019, showing the different periodic behaviour.
(a) The pulse sequence between 1700 and 1850~in mode B showing the low 
frequency periodic modulation seen in this mode. The phase variations 
corresponding to the periodic modulation feature (top window) are flat with 
close to zero slope across the profile. This highlights that the emission 
across the entire window changes simultaneously during periodic modulations.
(b) The average LRFS estimated in mode Q showing the subpulse drifting 
behaviour with $P_3=2.4P$. (c) The pulse between 15 and 185, in mode Q, showing
the presence of both subpulse drifting as well as periodic modulation. The 
phase behaviour corresponding to subpulse drifting in this pulsar is also flat
but limited to primarily the leading component.
\label{fig:B1758lrfs}}
\end{figure}

\begin{deluxetable}{cccccccccc}
\tablecaption{Estimating the periodic behaviour in PSR B1758-29
\label{tab:driftB1758}}
\tablewidth{0pt}
\tablehead{
   & \multicolumn{5}{c}{\underline{Subpulse Drifting}} &  & \multicolumn{3}{c}{\underline{Periodic Modulation}} \\
 \colhead{Frequency} & \colhead{$f_p$} & \colhead{$FWHM$} & \colhead{$P_3$} & \colhead{$\Delta\phi$} & \colhead{$d\psi/d\phi$} &   & \colhead{$f_p$} & \colhead{$FWHM$} & \colhead{$P_M$} \\
 \colhead{(MHz)} & \colhead{($cy/P$)} & \colhead{($cy/P$)} & \colhead{($P$)} & \colhead{(\degr)} & \colhead{(\degr/\degr)} &   & \colhead{($cy/P$)} & \colhead{($cy/P$)} & \colhead{($P$)}}
\startdata
 300-500 & 0.421$\pm$0.005 & 0.011 & 2.38$\pm$0.03 & -10.3\degr~: -7.0\degr & -0.4$\pm$0.9 &   & 0.017$\pm$0.004 & 0.009 & 59$\pm$13 \\
   &   &   &   &   &   &   &   &   &   \\
 550-750 & 0.410$\pm$0.004 & 0.010 & 2.44$\pm$0.03 & -9.0\degr~: -6.6\degr & 1.3$\pm$0.5 &   & 0.0145$\pm$0.0014 & 0.0033 & 69$\pm$7 \\
\enddata
\end{deluxetable}

\subsection{Subpulse Drifting and Periodic Modulation}
The presence of subpulse drifting in PSR B1758-29 was reported in the MSPES 
study of \citet{BMM16}, where periodic repetition with $P_3\sim2.5P$ was found
without any significant change in the drift phase across the pulse window. We 
have used the LRFS technique to measure the periodic variations of the wideband
observations at the two frequency ranges. The two emission modes show different
periodic behaviours in their pulse sequences. Fig.~\ref{fig:B1758lrfs}(a) shows
the LRFS corresponding to a pulse sequence in mode B and has a clear low 
frequency feature across the entire profile window, with relatively flat phase 
variations. This feature confirms the presence of periodic modulation in this 
pulsar that has not been reported in earlier studies. The short duration 
drifting feature is not visible in the LRFS suggesting the absence of 
systematic drifting during mode B, although the pulse sequence (see 
Fig.~\ref{fig:singlB1758}, left panel) might exhibit short bursts of periodic 
behaviour during this mode. Fig.~\ref{fig:B1758lrfs}(b) shows the average LRFS 
of mode Q from the 18 March, 2019 observations in the lower frequency band and 
shows the presence of prominent peak in the spectra around $f_p\sim0.4~cy/P$. 
This is primarily seen in the bright leading conal component with a flat phase 
variation across it, and is consistent with the subpulse drifting behaviour 
reported in the earlier work. Fig.~\ref{fig:B1758lrfs}(c) shows the LRFS of a 
pulse segment during Mode Q, with both subpulse drifting, seen as a high 
frequency peak, as well as periodic modulation as a low frequency feature, 
being present. The periodic modulation, that is usually a quasi-periodic 
behaviour, is seen across both modes of the pulsar, but systematic subpulse 
drifting is only present during the Q mode. 

The periodic modulation in pulsars is further divided into two categories, 
periodic nulling, where the single pulses transition between null and burst
states in a quasi-periodic manner, and periodic amplitude modulation, where the
transition happens between two emission states of varying intensity 
\citep{BMM20a}. There are around 30 pulsars that have periodic nulling in their
pulse sequence while around 20 with periodic amplitude modulation. Out of these
there are around 15 pulsars that have both period modulation and subpulse 
drifting in the same system. PSR B1758-29 is another addition to this select
group. In almost all of these cases periodic nulling and subpulse drifting 
coexist, with the only exception being PSR B1737+13 \citep{FR10}, also with an 
M type profile, where periodic amplitude modulation and subpulse drifting are
seen together. In case of PSR B1758-29 we were not able to clearly identify the
null pulses due to the presence of low intensity single pulses. It is more 
likely that the low frequency periodic modulation are a form of periodic 
amplitude modulation and not periodic nulling. In case of the Q mode the low 
frequency feature is most prominent when the short bursts of centrally bright 
pulses appear (see Fig.~\ref{fig:singlB1758}, right panel). These bursts 
typically repeat at intervals of 50-80$P$ that matches the periodic modulation
feature. Similarly, in case of the B mode there are regular intervals of bright
and weaker pulses, where the weak regions may have some nulls in them but also
exhibit low level emission. However, studies with higher sensitivity will be 
required to clearly identify the null pulses in both emission modes and find
conclusive evidence for the exact nature of periodic modulation in this pulsar. 

Table~\ref{tab:driftB1758} reports the measurement of the periodic behaviour in
the two frequency bands. This include $f_p$ and $FWHM$ for both the subpulse
drifting and periodic modulation, the periodicity of drifting, $P_3$, the 
longitude range of the leading component where the drifting behaviour is seen 
with sufficient sensitivity, $\Delta\phi$, as well as the slope of the drift 
phase over this range, $d\psi/d\phi$, and the periodicity of amplitude 
modulation, $P_M$. The estimated $P_3$ is around 2.4$P$ which matches earlier 
estimates and $d\psi/d\phi$ is close to zero at both frequencies, underlying 
the flat nature of the phase variations. The $P_M$ at both emission modes 
varies between 60-70$P$ and shows fluctuations with the feature becoming 
stronger at certain intervals and less prominent at other times.

\begin{figure}
\plottwo{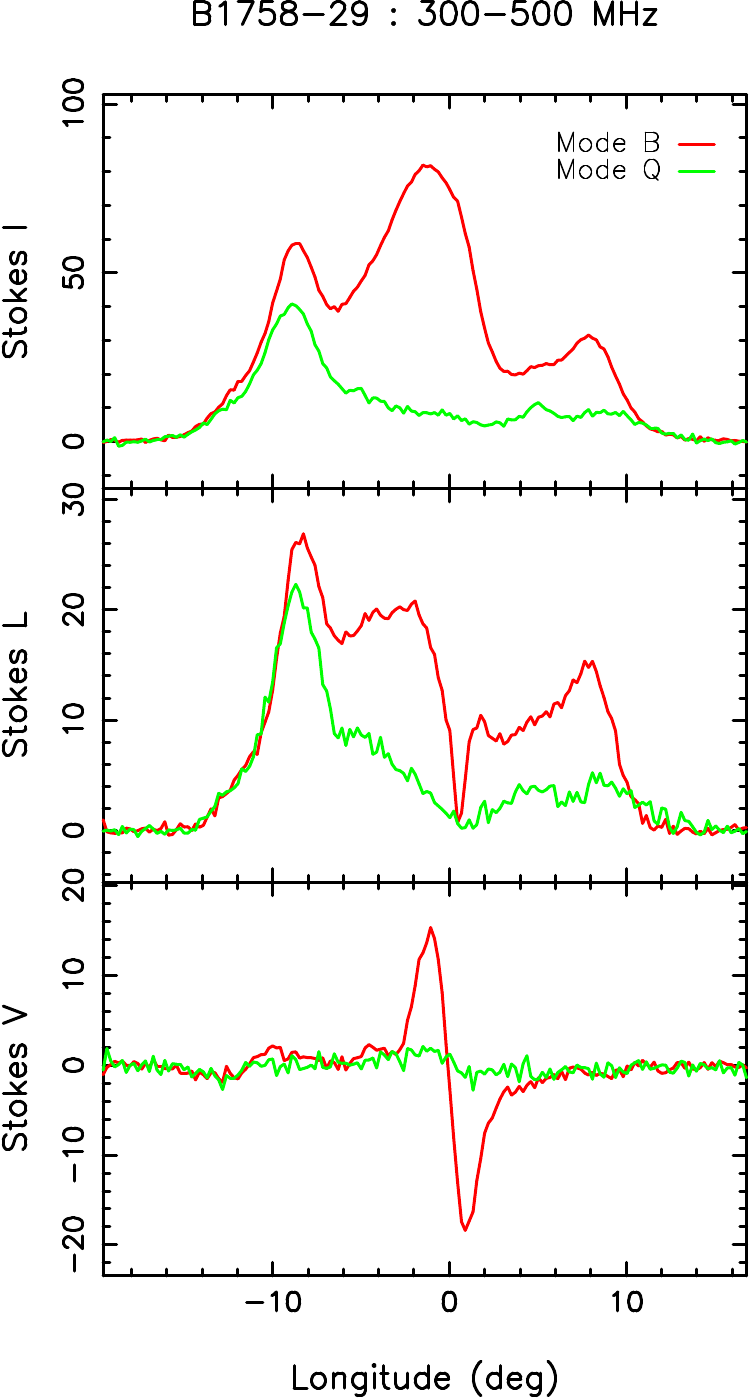}{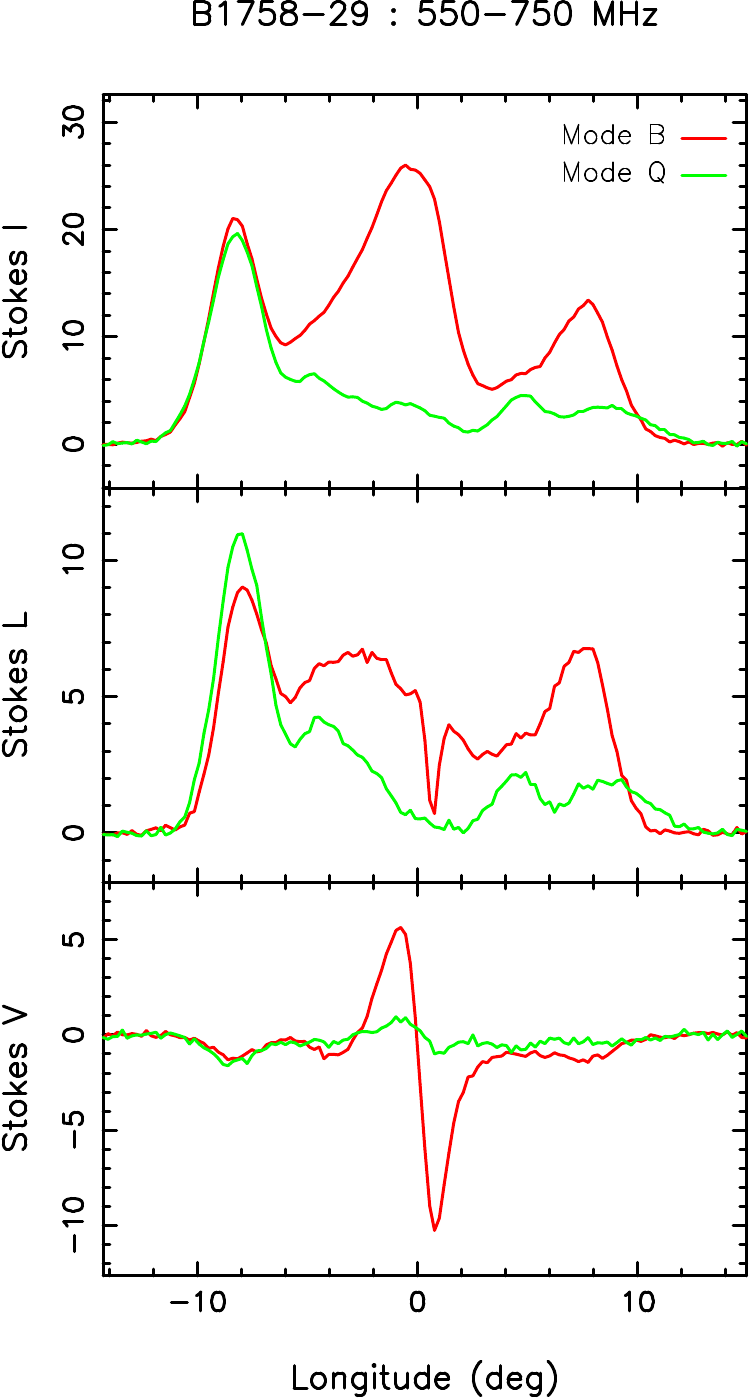}
\caption{The polarization behaviour across the average profiles of PSR
B1758-29, with the two modes B and Q shown in the same plot. The left panels
show the average profiles over the 300-500 MHz frequency range and the right
panels show the corresponding profiles averaged between 550-750 MHz. The top
window in each panel represents the total intensity profiles, the middle window
shows the variation of the linear polarization across the profile and the
bottom window the circular polarization behaviour.
\label{fig:polB1758}}
\end{figure}

\begin{figure}
\gridline{\fig{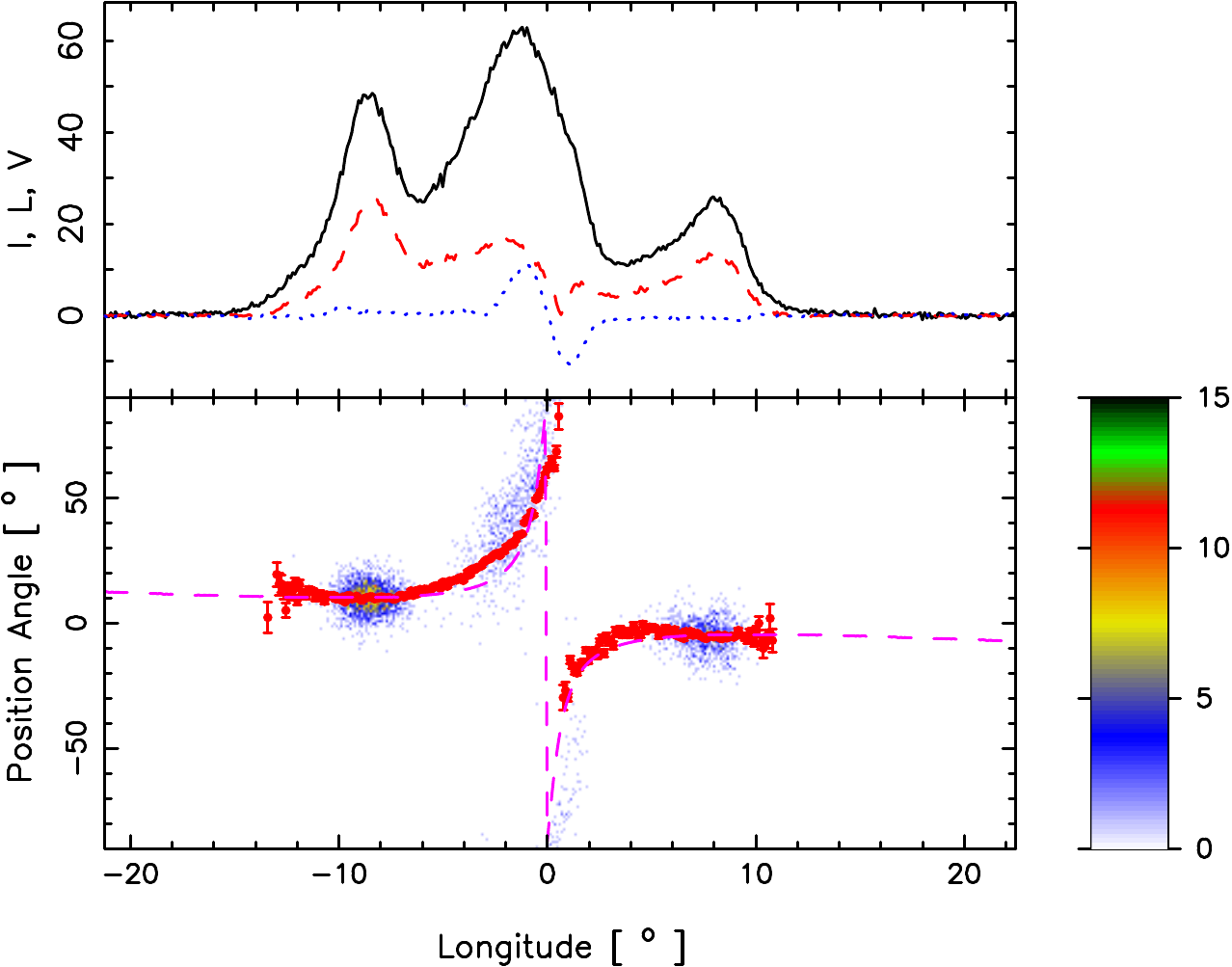}{0.45\textwidth}{(a) Mode B, 300-500 MHz}
          \fig{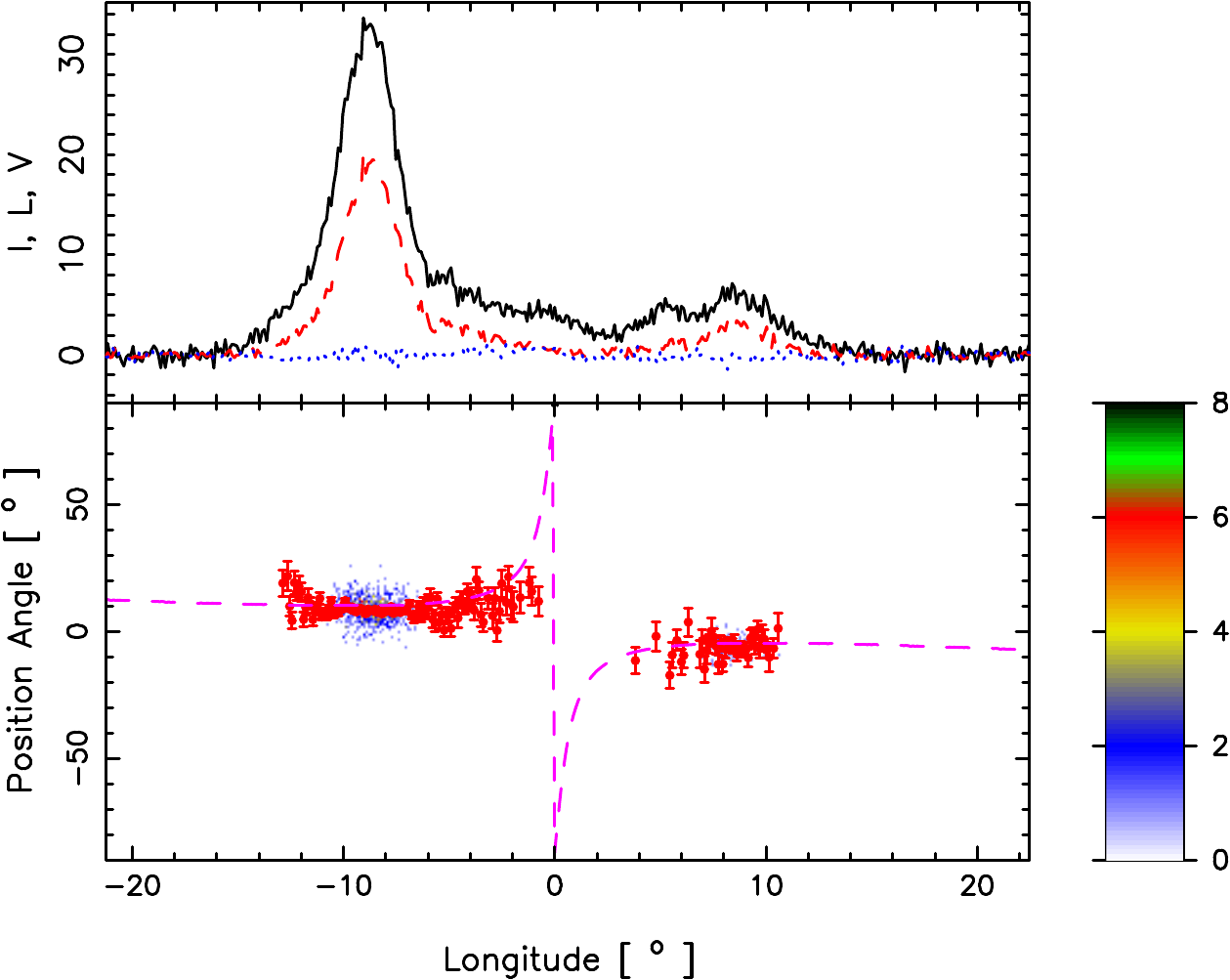}{0.45\textwidth}{(b) Mode Q, 300-500 MHz}
          }
\gridline{\fig{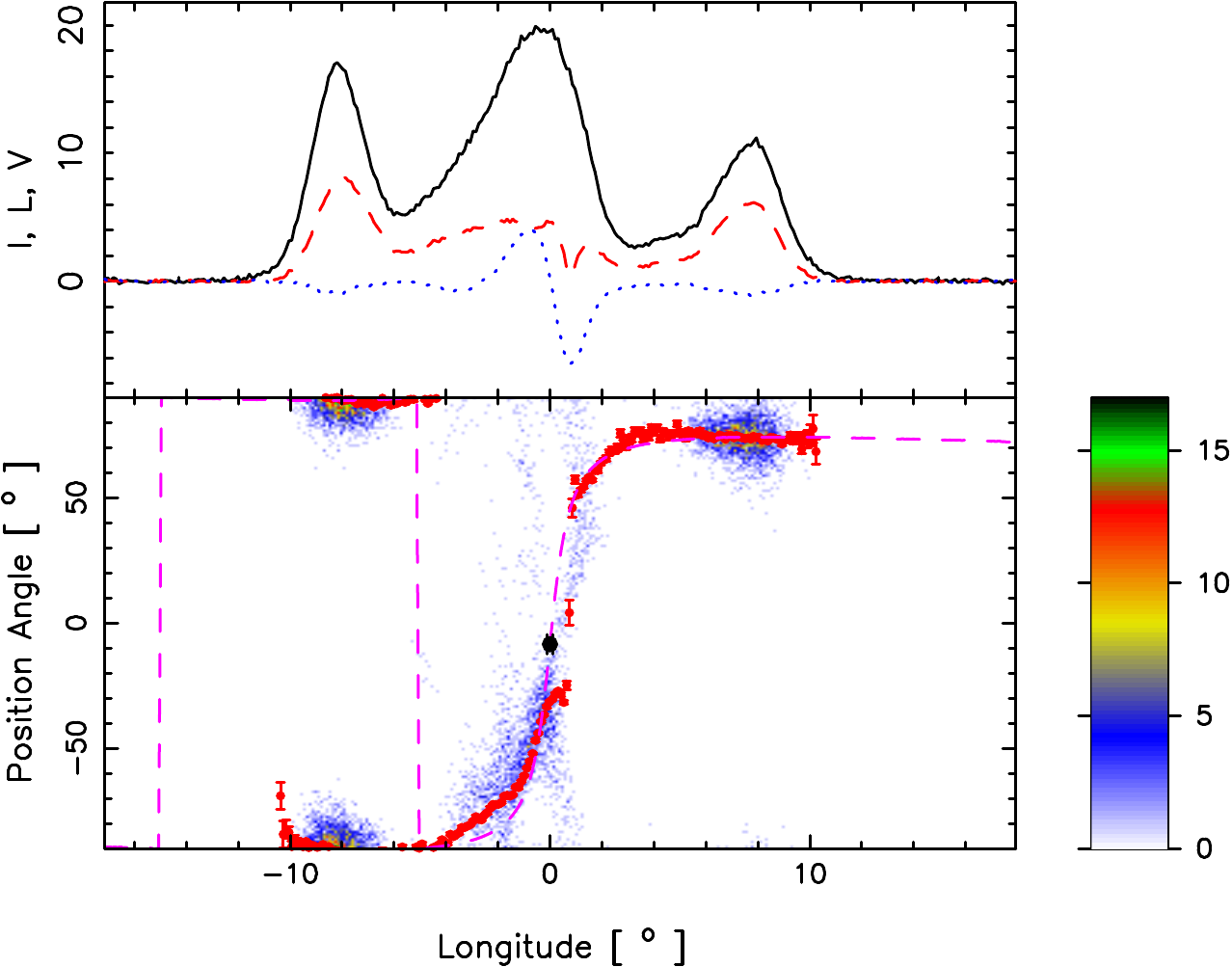}{0.45\textwidth}{(c) Mode B, 550-750 MHz}
          \fig{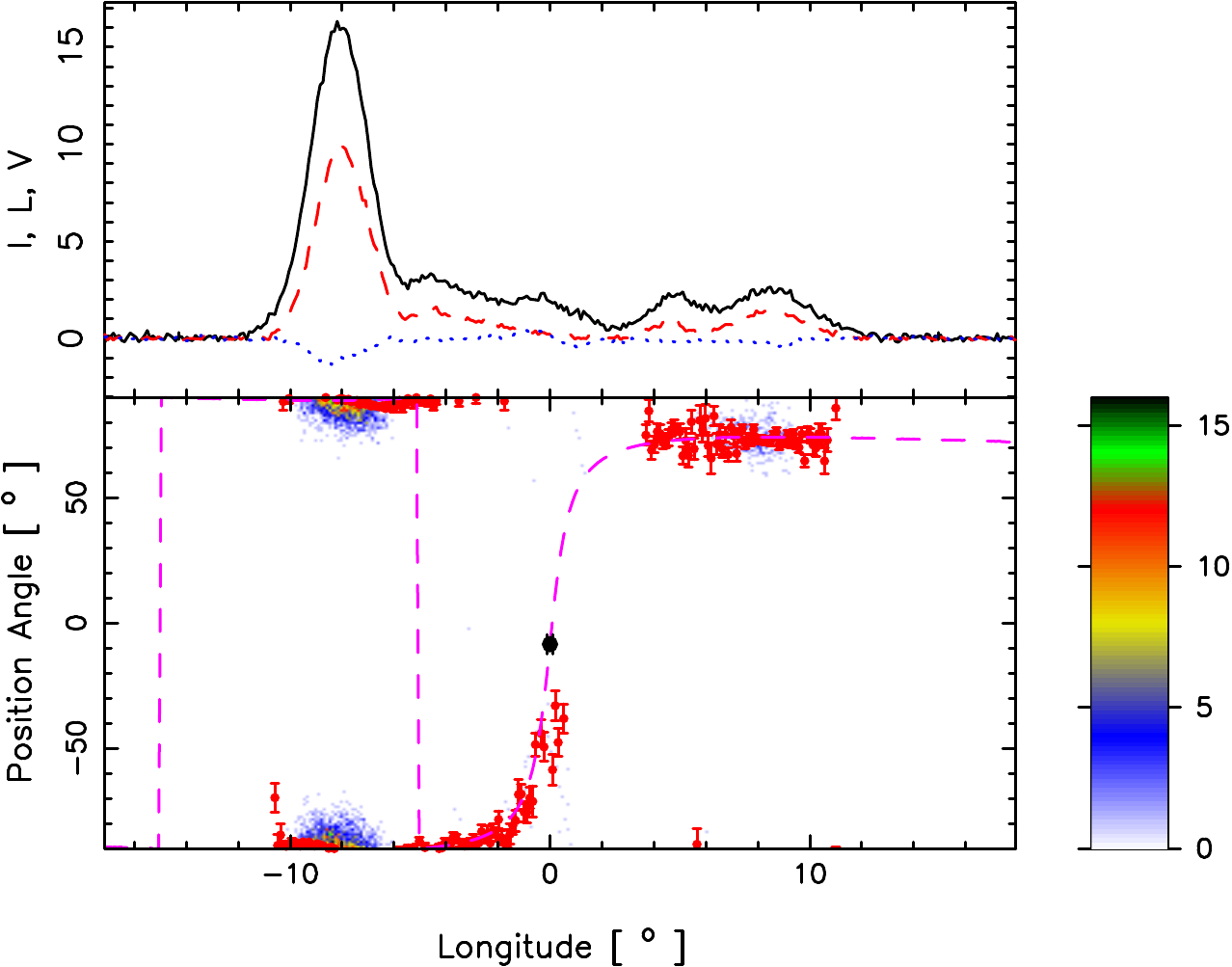}{0.45\textwidth}{(d) Mode Q, 550-750 MHz}
          }
\caption{The figure shows the polarization position angle (PPA) distribution of
the single pulse time samples of PSR B1758-29. The top windows (a) and (b)
shows the behaviour of mode B and Q, respectively, at the 300-500 MHz frequency
range. The bottom windows (c) and (d) shows the PPA of the two modes averaged
between 550 and 750 MHz. The top panel in each window shows the average profile
(black), the average linear (red) and circular polarization (blue) properties 
of the radio emission. In the bottom window the distribution of the single 
pulse PPA is shown as a colour scale, representing the number of points in each
location, as well as the average PPA behaviour across the window (red error 
bars). The rotating vector model (RVM) fits to the PPA is also shown in the 
figure (pink dashed line), along with the steepest gradient point of the RVM 
fit (black error bar).
\label{fig:B1758ppa}}
\end{figure}

\subsection{Polarization Properties and Emission height}
The average polarization behaviour of PSR B1758-29 is shown in 
Fig.~\ref{fig:polB1758}, for the 300-500 MHz frequency band (left panel) and 
the 550-750 MHz band (right panel). Both the B and the Q mode properties are 
superposed on the same plots showing the total intensity profiles (top window),
the linear polarization behaviour across the profile (middle window) and the
circular polarization (bottom window). The primary difference between the two
modes is encapsulated by the presence of the core component during mode B and 
its absence during the Q mode, and the associated depolarization at the center
of the profile in mode B as well as the sign changing circular polarization 
behaviour. The profiles of the two modes were aligned using a common reference
point estimated from the SG point of the PPA obtained from highly polarized 
time samples, to minimize the effects of polarization mode mixing 
\citep[see][for a detailed discussion]{MMB23a}. Similar to PSR B0844-35 we find
the emission region to remain identical in the two modes, both for the average 
profile and the individual components, with the exception of a slight flaring
seen in the trailing edge of the 550-750 MHz profiles of the Q mode. These 
results show that the location of the emission region remains unchanged during
the mode transition in PSR B1758-29.

Fig.~\ref{fig:B1758ppa} shows the PPA estimates from the significant time 
samples of the single pulses, both both modes and frequency bands. The average
PPA in all cases could be estimated using the RVM and the fits are shown in
the plots (dashed pink line) along with the SG point (black dot with error 
bars). Table \ref{tab:emhtB1758} reports the measurement of the different
quantities associated with the emission height estimates of modes B and Q at 
300-500 MHz and 550-750 MHz frequencies, including $\phi_o$, $\phi_l$, 
$\phi_t$, $\phi_c$, $\Delta\phi$ and $h_{\rm A/R}$. The estimated emission 
heights are around few hundred kilometers and agrees with the general trend. 
The height estimates of mode Q at 550-750 MHz are indeterminate because of the 
flaring in the trailing side of the profile which shifts $\phi_c$ towards 
$\phi_o$, thereby reducing $\Delta\phi$ (see eq. \ref{eq:hAR}).

\begin{deluxetable}{ccccccccc}
\tablecaption{Estimation of radio emission height in PSR B1758-29
\label{tab:emhtB1758}}
\tablewidth{0pt}
\tablehead{
 \colhead{Mode} & \colhead{Frequency} & \colhead{SG} & \colhead{$\phi_o$} & \colhead{$\phi_l$} & \colhead{$\phi_t$} & \colhead{$\phi_c$} & \colhead{$\Delta\phi$} & \colhead{$h_{\rm A/R}$} \\
    & \colhead{(MHz)} & \colhead{(\degr/\degr)} & \colhead{(\degr)} & \colhead{(\degr)} & \colhead{(\degr)} & \colhead{(\degr)} & \colhead{(km)}}
\startdata
  B & 300-500 & 90.2 & 0.35$\pm$0.15 & -15.45$\pm$0.22 & 13.11$\pm$0.22 & -1.17$\pm$0.31 & 1.52$\pm$0.34 & 342$\pm$77 \\
    & 550-750 & 102.3  & 0.38$\pm$0.09 & -12.32$\pm$0.22 & 11.89$\pm$0.22 & -0.22$\pm$0.31 & 0.60$\pm$0.32 & 135$\pm$72 \\
    &   &   &   &   &   &   &   &   \\
  Q & 300-500 & 118.0 & 0.92$\pm$0.15 & -15.23$\pm$0.22 & 12.46$\pm$0.22 & -1.39$\pm$0.31 & 2.31$\pm$0.34 & 520$\pm$77 \\
    & 550-750 & 119.9  & 0.60$\pm$0.40 & -12.32$\pm$0.22 & 12.54$\pm$0.22 & 0.11$\pm$0.31 & 0.49$\pm$0.51 & --- \\
\enddata
\end{deluxetable}

\section{Discussion} \label{sec:dis}
The analysis of the single pulse behaviour in PSR B0844-35 and PSR B1758-29 
have revealed the properties of the emission modes, with certain similarities 
that are seen on rare occasions in pulsars. Both pulsars have the presence of 
two distinct emission modes, with one mode showing increased intensity of the 
core component in the profile center. In both cases subpulse drifting is seen 
in the emission mode with the lower intensity core emission, with periodicity 
close to 2$P$. The single pulse emission becomes more chaotic with increased 
core activity and the well ordered drifting behaviour vanishes. Additionally, 
in the weaker state the core emission is usually faint, but shows short bursts 
of flaring from time to time. There are many similarities between the mode 
changing behaviour of these two pulsars and PSR B1237+25, that also has a M 
type profile \citep{SR05,SRM13}. The leading part of the profile in PSR 
B1237+25, including the core component, shows increased emission during the 
abnormal mode, but the central core emission is much weaker during the normal 
mode apart from instances of flaring. The pulsar also shows the presence of 
subpulse drifting with large phase variations similar to PSR B0844-35. Another
prominent example of M type profile associated with mode changing is PSR 
B2003-08 \citep{BPM19}, where the pulsar has four emission modes, modes A and B
with central core and showing subpulse drifting, mode D also with prominent 
core and showing periodic modulation in the form of periodic nulling, and an 
additional mode C with weaker core but the single pulses have periodic nulls. 
Our studies have now doubled this unique set of pulsars that exhibit mode 
changing and subpulse drifting in a system that has core component in the 
average profile.

In both modes of the two pulsars we have found profile widths to change with 
frequency (see Table~\ref{tab:profB0844} and \ref{tab:profB1758}), representing
the so called radius to frequency mapping, where the lower frequency profiles 
become progressively wider \citep{ET_MR02}. The radius to frequency mapping is 
a feature of the coherent curvature radiation along the diverging dipolar open 
magnetic field lines \citep{GLM04,MGM09}, where the characteristic frequency 
($\nu$) of emission is $\nu\sim\gamma^3/\rho_c$, here $\gamma$ is the Lorentz 
factor of the plasma responsible for the emission and $\rho_c$ is the radius of
curvature of the dipolar field lines. As the magnetic field lines diverge with 
distance from the neutron star surface, $\rho_c$ increases. If we consider 
$\gamma$ to be constant, i.e., the same plasma is responsible for all 
frequencies, then the lower frequencies are associated with higher values of 
$\rho_c$, and are emitted further away from the surface. If we assume the 
profile edges to be associated with the last open field lines then the radius 
to frequency mapping is a result of the lower frequency emission being emitted 
at higher heights due to the curvature radiation mechanism. 

We also found the location of the radio emission region in the two pulsars to 
remain largely unchanged during the mode transition. This result is consistent 
with previous reports of mode changing behaviour in several pulsars like, PSR 
B0329+54 \citep{BMR19}, PSR B1819-22 \citep{BM18b}, PSR B2003-08 \citep{BPM19},
and PSR B2319+60 \citep{RBMM21}, where the radio emission heights were also 
shown to be constant in the different modes. We detected short duration flaring
in the core region of the Q mode profile of PSR B1758-29, as well as the 
trailing edge of the profile, primarily at the higher frequency range. There 
are intriguing possibilities regarding the origin of such flaring, with the 
emission arising at slightly different heights in certain parts of the emission
beam, or possibility of partial illumination of the emission beam as seen 
prominently in the partial-cone pulsars \citep{MR11}. In the PSG model the 
properties of the outflowing plasma, including the Lorentz factor, is 
determined by the screening factor ($\eta$) in the IAR. It has been suggested 
that the mode changing arises as a result of changes introduced in the local 
magnetic field configuration of the IAR, due to perturbations introduced by 
Hall drift and thermoelectrically driven magnetic field oscillations 
\citep{GBM21}. These changes are likely to change the $\eta$ in the gap which 
in turn can result in different values of $\gamma$ for the two modes. Further 
exploration of these ideas, including flaring at different parts of the 
profile, would require rigorous works concerning the the nature of the 
outflowing plasma arising due to thermal regulation in the PSG.

The wideband observations have allowed a detailed characterisation of the 
subpulse drifting behaviour in the mode A of PSR B0844-35. The pulsar shows 
large phase variations across the different profile components, in the leading
outer cone as well as the trailing conal pair. These measurements have also
revealed the effect at the component boundaries, where jumps and reversals in 
the phase variations are seen. These effects seem to be different from the 
bi-drifting and phase switching behaviour seen in certain pulsars, where the
phase behaviour across the entire component show reversal or jumps compared to
other components and not just at the boundary between components \citep{CLM05,
W16,SvL17,BM18a,BMM19}. The most likely explanation for the boundary jumps in
the drifting phase is the overlapping of the subpulses belonging to two 
different drift sequences in the adjoining components, that are unconnected in
their phase behaviour. Similarly, flaring of the emission at the extreme edge
of the profile at certain intervals lead to additional reversals in the phase 
at the profile boundaries. The drift phase behaviour in the Q mode of PSR 
B1758-29 could only be measured in the leading component and shows flat 
behaviour across it, which is very different than PSR B0844-35. However, it is 
possible that the inner components may have large phase variations, like the 
drifting behaviour in PSR B2003-08 \citep[see Fig.~9 in][]{BPM19}, and would 
require more sensitive observations for detection. 

The complex phase behaviour in PSR B0844-35 suggests that surface magnetic 
field configuration is non-dipolar in nature. A prescription for modelling the 
non-dipolar magnetic field configuration of surface polar cap with an 
elliptical boundary has been proposed by \citet{BMM23}, where the complicated 
bi-drifting behaviour in PSR J1034-3224 has been reproduced from the sparking 
mechanism developing in a PSG \citep{MBM20,BMM20b}. Additional constraints to 
this model is provided by the frequency evolution of subpulse drifting measured
in this pulsar, where the phase variations across each component show steeper 
gradient with increase in frequency. This effect can once again be related to 
the radius to frequency mapping, where at higher frequencies the LOS traverses 
the outer region of the emission beam, and thereby samples a different 
trajectory across the evolution of the sparking pattern in the IAR. Thus in 
effect the measurement of the drifting features at multiple frequencies builds 
up a two dimensional map of sparking process in the IAR and shows the 
importance of multifrequency studies in understanding the drifting behaviour in
pulsars. A detailed modelling of the drifting behaviour in PSR B0844-35 using 
the PSG model would need to incorporate the various observational details 
uncovered in this work and requires a separate dedicated study. 

Data Availability : The polarized single pulse data for the two pulsars is
available for download at http://dataout.ia.uz.zgora.pl/ModeCent/

\section*{Acknowledgments}
We thank the referee for comments that helped to improve the paper. DM 
acknowledges the support of the Department of Atomic Energy, Government of 
India, under project no. 12-R\&D-TFR-5.02-0700. This work was supported by the 
grant 2020/37/B/ST9/02215 of the National Science Centre, Poland.

\bibliography{reflist}{}
\bibliographystyle{aasjournal}

\end{document}